\documentclass[twocolumn]{aastex631}
\usepackage{multirow}
\usepackage{diagbox}

\begin{document}

\title{Episodic Star Formation - I. Overview and Scatter of the Star-Forming Main Sequence}

\author[0009-0007-9365-9806]{Yuqian Gui}
\affiliation{Department of Astronomy, Tsinghua University, Beijing 100084, China}

\author{Dandan Xu}
\affiliation{Department of Astronomy, Tsinghua University, Beijing 100084, China}

\author[0009-0003-0166-0991]{Haoyi Wang}
\affiliation{Department of Astronomy, Tsinghua University, Beijing 100084, China}

\author[0009-0005-9427-8525]{Xuelun Mei}
\affiliation{Department of Astronomy, Tsinghua University, Beijing 100084, China}

\author[0000-0003-1588-9394]{Enci Wang}
\affiliation{School of Astronomy and Space Sciences, University of Science and Technology of China, Hefei, Anhui, 230026, China}

\author[0000-0002-8711-8970]{Cheng Li}
\affiliation{Department of Astronomy, Tsinghua University, Beijing 100084, China} 

\author[0000-0003-3735-1931]{Stijn Wuyts}
\affiliation{Department of Physics, University of Bath, Claverton Down, Bath BA2 7AY, UK}

\correspondingauthor{Yuqian Gui \& Dandan Xu}
\email{gyq24@mails.tsinghua.edu.cn}
\email{dandanxu@tsinghua.edu.cn}

\begin{abstract}

There is mounting evidence for episodic star formation cycles in both high- and low-redshift galaxies. This paper aims to understand the detailed physical processes behind such behaviors and investigate how such an episodic star-forming scenario can explain the scatter in star-formation rate (SFR) of star-forming main-sequence (SFMS) galaxies. This is achieved by tracing back in time the history of $z=0$ star-forming central galaxies in the TNG100 simulation over the past $7-8\,\rm Gyrs$.
As the first paper in this series, we provide an overview of the episodic star formation history.  
We find that two branches of star formation typically develop during each episode: while one branch happens in heavily metal-enriched gas in the centers of galaxies, a secondary branch starts in lower-metallicity regions at galaxy outskirts where fresh gas first arrives, and gradually progresses to the inner regions of galaxies. 
Additionally, the temporal variation in the SFR at galaxy outskirts is more significant than that at centers. As a consequence, the metallicities in both gas and young stars exhibit remarkably different distributions between SFR peaks and valleys. The resulting temporal SFR fluctuation within individual galaxies has an average of $\sim$0.2 dex, while the intrinsic differentiation between (the historical mean of) galaxies is $\sim$0.15 dex. These two together can well account for the scatter in SFR of $\sim$0.25 dex as observed for $z=0$ SFMS galaxies. 

\end{abstract}


\section{Introduction} 
\label{sec:intro}

The star formation history (SFH) of galaxies has always been among the most heated topics in the field of galaxy evolution. Observationally, the discoveries of galaxy bimodalities in color, morphology, and stellar kinematics (e.g., \citealt{ColorMagnitudeRelation_Tully1982,SDSSGalaxyColor_Strateva2001,Budavari2003,Hogg2003,Balogh2004,Bell2004,Baldry2004,Morselli2017,WangBitao2020,WangBitao2025}) have generally revealed two distinct populations of galaxies in the present-day Universe, one on the so-called `star formation main sequence' (SFMS, e.g., \citealt{SFGalaxy_lowz_Brinchmann2004,UV_SFR_Salim2007,SFMS_Noeske2007,SFMS_Daddi2007,SFRD_Elbaz2007,SFMS_Whitaker2012,SFMS_Speagle2014,SFMS_Pannella2015,SFMS_Pearson2018,VLACOSMOS_sSFR_Leslie2020,SFMS_Popesso2023}), also referred to as the `blue cloud,' and one with significantly lower or already ceased star formation activity, also dubbed the `red sequence' (e.g., \citealt{Kauffmann2003,QG_Faber2007,YingjiePeng2010,Schaye2010,YingjiePeng2015,Rodighiero2011,Renzini2016,Tacchella2022}). A classical picture for the formation path of the former suggests generally mild star-formation rate (SFR) variations throughout their evolutionary history as they ride along the SFMS to the present day. The latter population, on the other hand, underwent a transition from the SFMS at high redshift to a more quiescent or completely quenched status to the present day. This study focuses on the recent evolutionary history of the former class since $z\sim 1$, in particular, from perspectives regarding how galaxies move up and down the SFMS, the origin of its dispersion, and the properties of galaxies above and below the SFMS, which remain not fully understood.

The overall cosmic SFR density peaks around $z\sim 2$ and has been declining ever since \citep{Madau_and_Dickinson2014}. The specific star formation rate (${\rm sSFR}$) of SFMS galaxies also decreases strongly toward lower redshift as a consequence of a decreasing cosmological gas accretion rate (e.g., \citealt{SFMSEvo_Dutton2010, Dekel2013, SimonLilly2013, Tacchella2013, Forbes2014, Schreiber2015}). It has been observed that the SFMS has a typical scatter of 0.2--0.3 dex across a wide range of redshifts (e.g., \citealt{Noeske2007, Whitaker2012, Speagle2014, SFMS_Pearson2018}). The finite scatter in star-formation activity at fixed stellar mass is attributed to both internal and external factors. On the one hand, supernova and stellar wind feedback (e.g., \citealt{Dekel_and_Silk1986, Feedback_Murray2005, StellarFeedback_Ceverino_and_Klypin2009, FIRE_PhilipHopkins2014, FIRE3_PhilipHopkins2023, FIRE_El-Badry2016, Hayward_and_Hopkins2017, EunjinShin2023, Dome2025}) as well as supermassive black hole feedback (e.g., \citealt{DiMatteo2005, Feedback_Springel2005, Croton2006, Ciotti_and_Ostriker2007, Cattaneo2009, Fabian2012ARAA, Cicone2014, Cicone2016}) may affect both the thermal and kinematic state of the interstellar medium (ISM) or even the circumgalactic medium (CGM), and thus regulate the rate of star formation. On the other hand, galaxy interaction and merger activities may affect cold gas accretion rates and thus modulate galaxy SFRs (e.g., \citealt{Merger_Barnes_and_Hernquist1991, Mihos_and_Hernquist1996, Merger_and_Accretion_Conselice2003, Cox2006, Merger_PhilipHopkins2006, Smethurst2015, SenWang_2022}). As a consequence, the SFR of individual galaxies can fluctuate on both short and long timescales, in particular among low-mass or high-$z$ galaxies, resulting in the observed scatter around the SFMS (\citealt{White_and_Rees1978, SFMSEvo_Dutton2010, Dekel_and_Burkert2014, Somerville_and_Dave2015ARAA, Feldmann2017, Orr2017, Sparre2017, Highz_Tacchella2018, Behroozi2019}; \citealt{Caplar_and_Tacchella2019, Iyer2020, JennyWan2024, Mintz2025, THESANZOOM_McClymont2025}).

In the high-$z$ Universe ($z>1-2$), the fluctuation in SFR is also highly relevant in explaining the discovery of overabundant UV-bright galaxies as recently revealed by JWST (\citealt{UVBrightJWST_DonnanCT2023, DonnanCT2024, UVG_JWST10_Leethochawalit2023L, PerezGonzalez2023, BurstyJADES_Endsley2024, Robertson2024, Whitler2025}). This is because a bursty/fluctuating SFH with a high variability in SFR may temporarily scatter high-$z$ galaxies to large UV luminosities and thus explain such observations, along with the scatter of the SFMS (\citealt{AnglesAlcazar2017, FaucherGiguere2018, Tacchella2020, UV_Mason2023, XuejianShen2023, GuochaoSun2023,  Kravtsov_and_Belokurov2024, UVBurst_Endsley2024, Ciesla2024, THESANZOOM_McClymont2025, Mintz2025}).

Fluctuating/bursty SFHs can be considered one manifestation of episodic behavior in galaxy growth. High-$z$ galaxies typically go through episodic cycles of gas compaction, depletion, and replenishment, and eventually cease their star formation and become quenched (e.g., \citealt{Quenching_Zolotov2015, SFGGas_Tacchella2016, Tacchella2020, THESANZOOM_McClymont2025}). It is worth noting that such a bursty/fluctuating SFH is not unique for high-$z$ galaxies, and more and more observational evidence also suggests their universality among star-forming galaxies across all redshifts, in particular, among lower mass galaxies (\citealt{Tolstoy2009ARAA, McQuinn2010a, McQuinn2010b, Weisz2011, EnciWang_and_Lilly2020a, JunYin2023, 2025MunozLopez, MarissaPerry2025, JennyWan2025}). Unlike (the present-day {\it quiescent} descendants of) their high-$z$ counterparts, these galaxies have nearly always maintained their SFRs to a certain level and never evolved to a long-term quenched status, and therefore remain star-forming to the present day.

In this regard, a number of theoretical studies have conducted relevant investigations (e.g., \citealt{SimonLilly2013, EnciWang2019, EnciWang_2021, EnciWang_and_Lilly2022a, EnciWang_and_Lilly2022b}). Based on the variation in star formation across the population of the MaNGA sample, \cite{EnciWang2019} proposed that the scatter of the SFMS can be understood by the time-varying gas inflow rate. This scenario predicts that the scatter of galaxies on the SFMS strongly depends on galaxy size, with smaller sizes for larger scatter. Interestingly, this is seen in observations across different datasets and redshifts (\citealt{StijnWuyts2011, WangKongPan2018}).

Such a fluctuating SFH in lower-$z$ star-forming galaxies has also been found in many of the latest cosmological hydrodynamical simulations. For example, utilizing the FIRE simulation \citep{FIRE_Hopkins2014, FIRE2_Hopkins2018, FIREbox_Feldmann2023}, \cite{Cenci2024} investigated the `breathing mode' of galaxy starbursts on long and short timescales. With the Magneticum Pathfinder simulation \citep{Hirschmann2014}, \cite{Fortune2025} revealed cycles of quenching and rejuvenation on timescales of $\sim 1$ Gyr for a large fraction of $z=0$ SFMS galaxies. \cite{SenWang_2022} also demonstrated a clear episodic SFH of present-day star-forming disk galaxies using the TNG100 simulation (\citealt{TNGCite04, TNGCite03, TNGCite05, TNGCite01, TNGCite02}). In this particular study, the authors presented how the CGM of (central) galaxies acquires a supply of cold gas ($T<2\times10^4\,{\rm K}$) and inherits angular momentum as a consequence of galaxy mergers and interactions at $\sim 100$ kpc scales (see also \citealt{WangSen2025}). At the same time, the CGM also receives thermal and kinetic modulation of internal feedback processes out to several tens of kiloparsecs. The combined effect regulates the replenishment and depletion of the cold gas reservoir in a rhythmic fashion, leading to a pulsating motion of the cold circumgalactic gas highly synchronized with an episodic SFH of the galaxy.

This series of papers is a follow-up of \cite{SenWang_2022}, conducting a detailed investigation to better understand the episodic star formation behavior of present-day SFMS galaxies and how such a behavior may explain some of the observed relations. This is achieved through tracing back in time the history of star-forming galaxies at $z=0$ in the TNG100 simulation and investigating the coevolution of various galaxy properties as galaxies go through SFR peaks and troughs in their history since $z\sim 1$. In this first paper of the series, we present an overview and the timeline of the episodic SFH of $z=0$ SFMS galaxies. In particular, we show that major differences exist among galaxy properties between SFR peaks and valleys, and that the resulting temporal SFR fluctuation within individual galaxies ($\sim$0.2 dex) together with intrinsic scatter among different galaxies ($\sim$0.15 dex) can account for the scatter of the SFMS today ($\sim$0.25 dex). In the second paper of this series (X. Mei et al. in preparation), we will show that the simulation successfully reproduces the observed anticorrelation between SFR and metallicity, and suggest that the episodic SFH can also offer a plausible explanation for the observed fundamental metallicity relation \citep{Mannucci2010}. In the third paper of this series (H. Wang et al. in preparation), we further present the key role that cold gas plays under the combined internal and external modulation across a wide range of scales in such an episodic formation scenario.

This paper is organized as follows: In Section \ref{sec:method}, we introduce the details of the simulation, galaxy sample selection, and galaxy properties investigated in this study. In Section \ref{sec:results-overview}, we present an overview of episodic SFH by investigating the evolution of gas and stellar mass, metallicity, and their spatial distributions. We will show that two star-formation branches of distinct chemical signatures exist during each episode, and the metallicities in both gas and young stars exhibit remarkably different distributions between SFR peaks and valleys.
In Section \ref{sec:results-scatter}, we present the degree to which the temporal fluctuation in SFR can account for the scatter of SFMS. The conclusions and some further discussions are given in Section \ref{sec:conclusions}. Throughout the paper, we adopt the same cosmology as used in the IllustrisTNG simulation, which is based on the Planck results (\citealt{Planck_Collaboration2016}), i.e., a total matter density of $\Omega_{\rm m} = 0.3089$, a baryonic matter density of $\Omega_{\rm b} = 0.0486$, and a Hubble constant $h = H_0/(100\,{\rm km s}^{-1} {\rm Mpc^{-1}}) = 0.6774$, assuming a flat $\Lambda$CDM Universe.

\section{Methodology} \label{sec:method}

\subsection{The simulation}\label{sec:method_TNG100}
The \textit{Next Generation Illustris Simulations} (IllustrisTNG, hereafter TNG; \citealt{ TNGCite02, TNGCite03, TNG50_02, TNGCite04, TNGCite05, TNGCite01, TNG50_01}) are a set of state-of-the-art magnetohydrodynamic cosmological simulations using the moving-mesh code \texttt{AREPO} \citep{AREPO} for galaxy formation and evolution. The simulation, compared to previous galaxy formation cosmological simulations, has implemented various modifications to galactic wind feedback, stellar evolution, and chemical enrichment \citep{TNGmodel}. In this study, we use one of the TNG versions with a cubic box of $110.7\ \rm Mpc$ side length (hereafter TNG100), which was run with a gravitational softening length of 0.5 $h^{-1}\rm kpc$ and mass resolution of $1.4\times10^6\,\rm M_\odot$ and $7.5\times10^6\,\rm M_\odot$ for the baryonic and dark matter, respectively. In TNG100, the \texttt{SUBFIND} algorithm \citep{subfind_1,subfind_2} is used to identify host dark matter halos. General galaxy properties were calculated and publicly released by the TNG Collaboration\footnote{\url{http://www.tng-project.org/data/}} \citep{TNGDataRelease}.

\subsection{Sample selection: evaluating scatter in SFMS}
\label{sec:method_galaxysample}

One goal of this study is to demonstrate to what extent the $z=0$ SFMS scatter can be accounted for by the temporal fluctuation in SFR within individual star-forming galaxies according to the TNG100 simulation. To first form our SFMS sample, we take all star-forming central galaxies with stellar mass $M_\ast \geqslant 5\times10^{9}\,\mathrm{M_\odot}$ at $z=0$ from the simulation, explicitly avoiding all satellites whose SFH may be affected by their host group or cluster environment (\citealt{YingjiePeng2010, Bahe_and_McCarthy2015}). Here `star-forming or SF' is defined as the logarithmic specific ${\rm SFR}$ $\log\, [{\rm sSFR}/{{\rm Gyr}^{-1}}] \geqslant -1.5$, which is approximately 0.5 dex below the mean ${\rm sSFR}$ of galaxies with $M_\ast < 3\times10^{10}\,\mathrm{M_\odot}$ at $z=0$ (e.g., \citealt{Genel2018, Donnari2019, ShengdongLu_2021a, ShengdongLu_2021b}). Under this definition, our SFMS sample contains 3644 galaxies.

We further trace the merger history of these galaxies and specifically consider the main progenitor branch up to $z= 1$ (snapshot 050). At each redshift, we determine the ridge of SFMS using a linear regression fitted to the $\log {\rm SFR}-\log M_\ast$ distribution of the collection of the corresponding galaxy progenitors. We then calculate for each galaxy the difference in the logarithmic ${\rm SFR}$ from the ridge at that redshift (i.e., the main-sequence offset), denoted as $\Delta\rm MS(z)$.  
For the derived $\Delta\rm MS$ among all sample galaxies at all redshifts, we capture the degree of main-sequence offset for each galaxy using two different statistics: (1) the present-day main-sequence offset ($\Delta\rm MS_{z=0}$); and (2) the historically averaged main-sequence offset since $z=1$ ($\text{avg}(\Delta\rm MS)$), denoting in general whether a galaxy has primarily lived above/below the main-sequence ridge, and by how much on average. In order to also capture the temporal ${\rm SFR}$ fluctuation within a given galaxy since $z= 1$, we fit a linear relation to the $\log{\rm SFR}-\log(1+z)$ path of each galaxy. We then calculate, for a given galaxy, the scatter $\sigma_{\rm SFR, rel}$ in logarithmic ${\rm SFR}$ relative to its long-term redshift evolution. Specifically, $\sigma_{\rm SFR, rel}=\sqrt{\frac{1}{N}\Sigma[\log{\rm SFR}(z) - \overline{\log \rm SFR_{\mathrm{lin}}}(z)]^2}$, where summation is carried out over all $N$ snapshots since $z=1$ and $\overline{\log\rm SFR_{\mathrm{lin}}}(z)$ is the long-term redshift-dependent value given by the linear fitting. For all three properties introduced above, we quantify their 16th, 50th, and 84th percentiles over all galaxies in the SFMS sample. The results are presented in Section \ref{sec:results-scatter}.

We also note that the reason we only trace the main progenitor history back to $z= 1$ is to focus on the relatively secular evolutionary phase of present-day star-forming galaxies. During this phase, galaxies are expected to be in a quasi-steady state, with self-regulated star formation governed by regular gas accretion and feedback processes.

\subsection{A further refined sample of typical episodic star-forming galaxies}\label{sec:method_SelESFG}

\begin{figure}[t!]
    \centering
    \includegraphics[width=\linewidth]{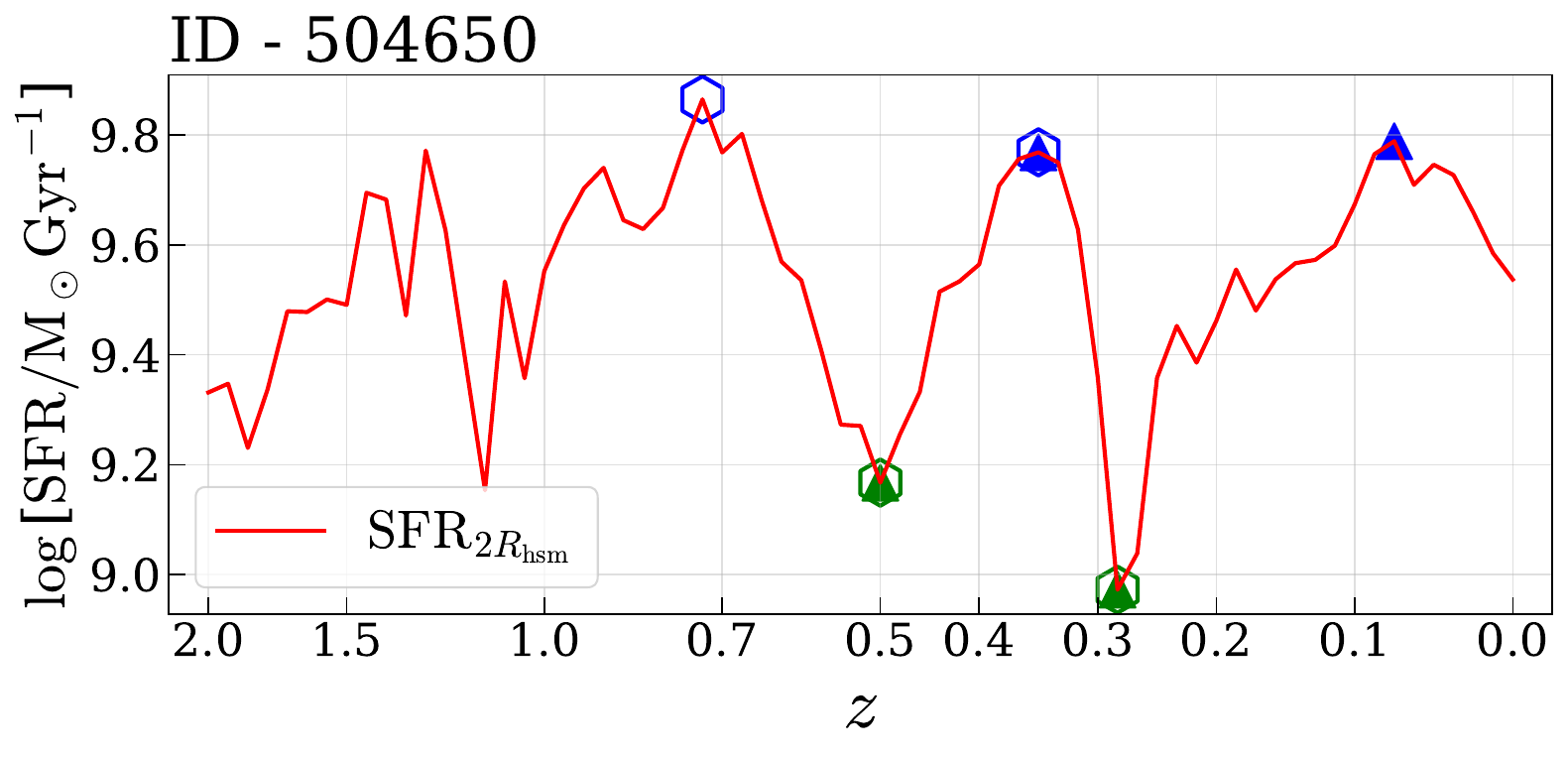}
    \caption{The SFH of one of our sample galaxies (ID at $z=0$ is 504650). The red line represents the SFR of this galaxy evaluated within $2R_{\mathrm{hsm}}$. The green and blue upper triangles represent the valleys and peaks in the upward pairs. The blue and green hollow hexagons represent peaks and valleys in the downward pairs.}
    \label{fig:episodicSFH_example}
\end{figure} 

In order to demonstrate typical episodic star formation behavior and conduct statistics at SFR peaks and valleys for key galaxy properties, we further select a refined sample from all present-day star-forming central galaxies, requiring that typical ``episodic'' galaxies shall have at least two valley-to-peak (upward) pairs or two peak-to-valley (downward) pairs since $z=1$, as an indication of experiencing star formation episodes (see Appendix \ref{sec:appendix_a} for a detailed description of the pair searching algorithm adopted). Below, we explain the detailed criteria and procedure.

For the full sample of 3644 present-day star-forming central galaxies, we carry out a close inspection of the redshift evolution along the main progenitor branch of each galaxy. The number of galaxies that exhibit at least two peak-valley pairs in their history since $z=1$ depends on $\Delta\log\,[\mathrm{SFR}/\mathrm{M_\odot Gyr^{-1}}]$ -- the difference in ${\rm SFR}$ between the identified adjacent peak and valley in a given pair. If requiring $\Delta\log\,[\mathrm{SFR}/\mathrm{M_\odot Gyr^{-1}}]\geqslant 0.3$ dex, then nearly all (98\%) galaxies satisfy the criteria of possessing at least two such upward or downward pairs. This fraction drops to 70\% if $\Delta\log\,[\mathrm{SFR}/\mathrm{M_\odot Gyr^{-1}}]\geqslant 0.5$ dex (see Fig.\,\ref{fig:peakvalleynumber} in Appendix \ref{sec:appendix_threshold}).

In order to gain insight into the most characteristic features of episodic SFH, we have adopted 0.5 dex (in comparison to an $\sim 0.25$ dex scatter in the $z=0$ SFMS) as the $\Delta\log\,[\mathrm{SFR}/\mathrm{M_\odot Gyr^{-1}}]$ {\it threshold} for this study. This yields a total of 2536 galaxies from the total of 3644 $z=0$ star-forming central galaxies, with a total of 5003 upward pairs and 5128 downward pairs since $z=1$. Fig.\,\ref{fig:episodicSFH_example} shows the SFH of an example galaxy (subfindID-504650 at $z=0$, ID-504650 in short), where the identified peaks and valleys are labeled by blue and green symbols, and the upward pairs and downward pairs are denoted by solid triangles and hollow hexagons, respectively.

Due to the computational cost of the subsequent calculation on peak-valley statistics for galaxy stellar and cold gas properties, we further reduce the sample size by {\it randomly} selecting a subsample of $\sim400$ galaxies from the parent sample of 2536 galaxies from above. Specifically, we consider two mass ranges based on galaxy stellar mass within $2R_{\rm hsm}$, i.e., $9.5 \leqslant \log M_{\ast,2R_{\rm hsm}}/\mathrm{M_\odot} \leqslant 10.3$ and $10.3 < \log M_{\ast,2R_{\rm hsm}}/\mathrm{M_\odot} \leqslant 11.2$, in order to investigate possible mass dependence. This results in a total of 408 galaxies, with 201 galaxies in the lower-mass range and 207 galaxies in the higher-mass range. A further comparison between this refined galaxy sample (408), the total sample with typical episodic star-formation behavior (2536), and the full sample of $z=0$ star-forming central galaxies (3644) is presented in Fig.\,\ref{fig:samplecomparison} in Appendix \ref{sec:appendix_samplecomparison}, where the distributions of central stellar mass $\log M_{\ast,2R_{\rm hsm}}$, SFR, present-day main-sequence offset $\Delta{\rm MS}_{z=0}$, and relative SFR fluctuation $\sigma_{\rm SFR, rel}$ among the three populations are presented. As can be seen, the final subsample of 408 galaxies is an unbiased representation of the total star-forming central galaxy population.

\subsection{Galaxy properties studied}\label{sec:method_properties}

In this work, we investigate the evolution of several galaxy properties along with SFR, including $g-r$ color, stellar half-mass radius $R_{\mathrm{hsm}}$, cold gas mass $M_{\rm cold}$, stellar and gas-phase metallicity $Z_\ast$ and $Z_{\rm gas}$. In particular, the properties are usually measured within $2R_{\mathrm{hsm}}$, which are referred to as global galactic quantities.

For the gas properties, we follow two components of cold gas.
The first component is {\it non-SF} gas with an effective temperature $T < 2 \times 10^4\,\mathrm{K}$ -- a typical temperature below which atomic hydrogen can cool and thus is able to form stars  (provided it reaches sufficient density) (\citealt{Keres2005, Dekel2006, Dekel2009Nature}). Observationally, this gas can be traced by the Ly$\alpha$ line (\citealt{KatzGunn1991, Fardal2001ColdGasLyA}) and HI 21 cm emission (\citealt{Draine2011book, SaintongeCatinella2022}). This cold but {\it non-SF} gas sits at a density below the adopted star-forming threshold of $n_{\rm H} \sim 0.1\,{\rm cm^{-3}}$ (by the simulation), representing the radiatively cooled, dense gas in galaxies (\citealt{Vogelsberger2012, Martizzi2019}). We refer to this gas component as the `cold non-SF gas reservoir.' The second component is {\it star-forming} gas. This gas sits at a density above the star-forming threshold, and is numerically enforced with an effective equation of state which describes a two-phase medium composed of both hot and cold gas of the interstellar medium \citep{Springel_and_Hernquist2003}. We refer to this gas as the ‘SF gas pool.’ 

While the majority of {\it star-forming} gas resides in inner regions of galaxies ($<2R_{\rm hsm}$), the cold {\it non-SF} gas is predominantly located at larger radii. The increase of the {\it non-SF} cold gas reservoir can originate from fresh accretion externally, as well as from cooling of the hotter-phase gas within the galaxy halo. The {\it star-forming} gas pool arises from compaction of {\it non-SF} cold gas that lies just below the density threshold. Regarding the possible fates of gases in these two phases (in central galaxies), the {\it non-SF} cold gas reservoir in general has two key outcomes: (1) compaction into {\it star-forming} gas, and (2) being heated above the temperature limit or expelled from the evaluated aperture by feedback. For {\it star-forming} gas, star formation (conversion into stars) and feedback serve as two primary outlets for the direct {\it star-forming} gas pool (see \citealt{Martizzi2019} for more details on distributions of gases in different phases).

The combination of these two gas components constitutes the total cold gas budget. In this paper, we track the evolution of each component individually as well as their combined behavior. {\it Non-SF} cold gas serves as a reservoir for a potential future {\it star-forming} gas, while {\it star-forming} gas acts as an immediate pool for ongoing star formation. Specifically, we follow the evolution of the enclosed mass, surface density, and metallicity distribution of both types of cold gas components during the episodic SFH of galaxies.

\subsection{Stacking galaxies for peak-valley statistics}\label{sec:method_peakvalley}
In order to demonstrate the most significant features and differences between the peak and valley moments, we stack galaxy properties at the identified peaks and valleys of all galaxies in the refined galaxy sample. In order to further extract the general statistical trends around peaks and valleys, we take the neighboring snapshots around these moments according to a given time offset. To reduce noise/fluctuation at shorter time scales, for any given galaxy property at a specific snapshot $i$, we first calculate the mean of this property among the three adjacent snapshots, i.e., $i-1$, $i$, and $i+1$. For any given time offset around peaks/valleys, the corresponding galaxy properties are then interpolated using the smoothed values at two adjacent snapshots. We line up the moments of all peaks/valleys and take the smoothed and interpolated values of a given galaxy property under investigation as its zero-point. For any given time offset, we then calculate the difference in this property from the zero-point value. By doing so, the galaxy-to-galaxy difference among the population can be markedly eliminated, only preserving the internal and temporal offsets relative to peaks and valleys for any given galaxy. Then we can investigate systematic differences between the peak and valley moments as well as the evolutionary trend of galaxy properties around these extrema.

\section{Results: Overview}\label{sec:results-overview}

In order to investigate characteristic galaxy properties at SFR peaks and valleys, we selected typical episodic star-forming galaxies that exhibit at least two upward or downward pairs with peak-valley SFR difference $\Delta\log\,[\mathrm{SFR}/\mathrm{M_\odot Gyr^{-1}}]\geqslant 0.5$ dex (see Section \ref{sec:method_SelESFG} for details). We note that the average number of peak–valley pairs per galaxy and the time interval between them depend on $\Delta\log\,[\mathrm{SFR}/\mathrm{M_\odot Gyr^{-1}}]$. In general, larger fluctuation amplitudes yield fewer peak–valley pairs and longer timescales, while smaller amplitudes produce more pairs with shorter timescales. As shown in Figs.\,\ref{fig:peakvalleynumber} and \ref{fig:timeintervals} in Appendix \ref{sec:appendix_threshold}, for $\Delta\log\,[\mathrm{SFR}/\mathrm{M_\odot Gyr^{-1}}]\geqslant 0.3$ dex, nearly 98\% of the 3644 star-forming central galaxies exhibit, on average, three peak–valley pairs since $z=1$, with a ypical peak-valley time interval around 600 Myr. When the threshold increases to 0.5 dex, 70\% of the full sample (i.e., 2536 galaxies) show an average of two peak–valley pairs, with an average time interval of $\sim$1.25 Gyr. We note that such timescales are approximately an order of magnitude shorter than the Hubble time, and roughly comparable or just a few times the galaxy orbital timescale at 2$R_{\rm hsm}$, which very likely indicates that such episodic star formation variations are more related to galaxy-scale dynamical processes.

In this section, we first present an overview of the episodic star formation scenario, including a detailed pattern regarding how the star formation zone propagates through a galaxy during each episode, as well as a detailed comparison of galaxy properties at SFR peaks and valleys (using a randomly selected but further downsized sample of $\sim 400$ galaxies to reduce computational cost, see Section \ref{sec:method_SelESFG}). In Sections \ref{sec:result_color} and \ref{sec:result_coldgas}, we show how galaxy color and the cold gas content vary with time, respectively. In Section \ref{sec:result_2branches}, we demonstrate the presence of two star-formation branches during each episode, and in Section \ref{sec:result_diffmet}, we show how they are intertwined with different radial distributions in metallicity of young stars and gas between SFR peak and valley moments.

\subsection{Temporal variation in galaxy color}\label{sec:result_color}

\begin{figure}[t!]
    \centering
	\includegraphics[width=\columnwidth]{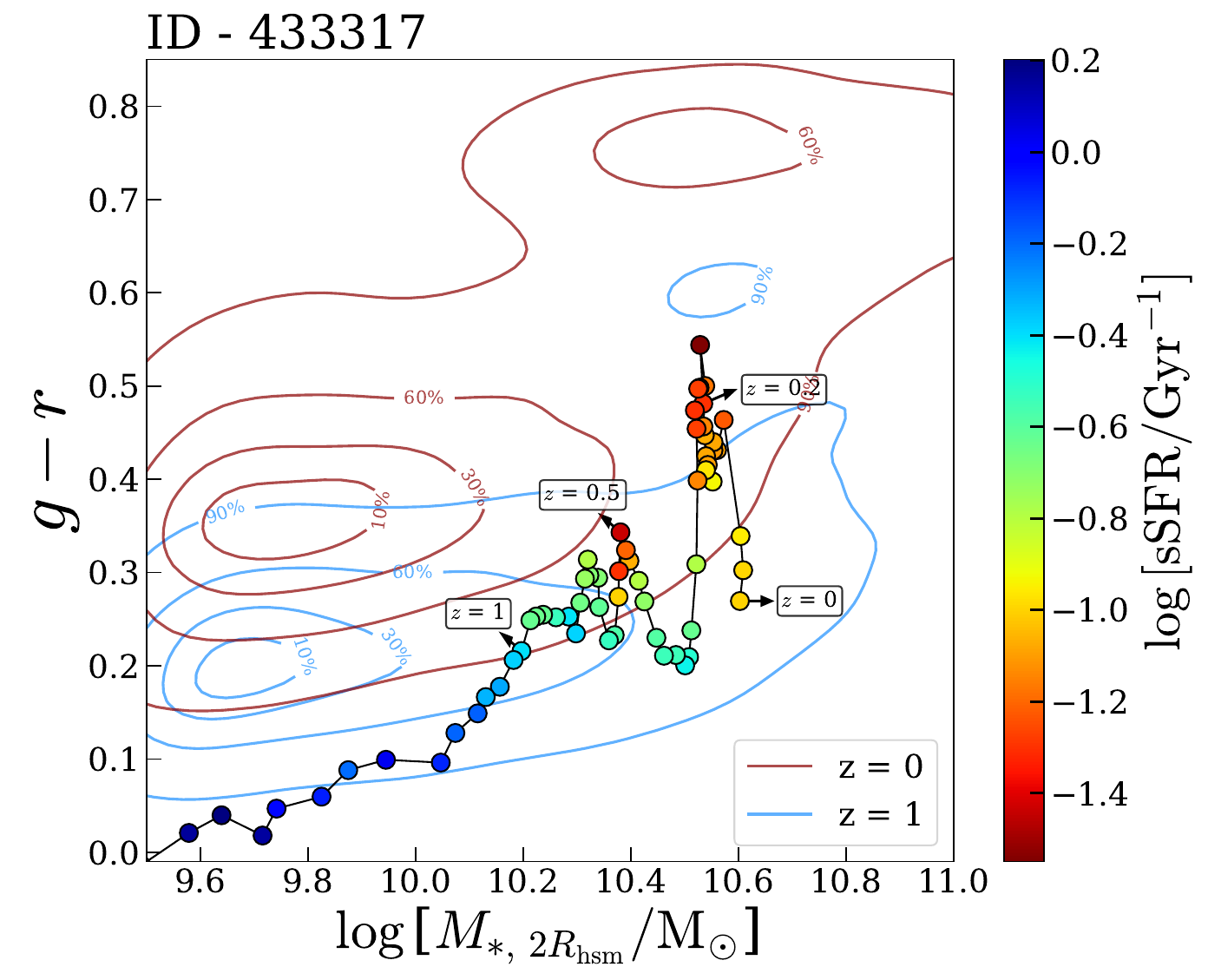}
    \caption{The $g-r$ color vs. log $M_\ast$ evolutionary track for an example galaxy (ID-433317). The contours represent the color distribution as a function of stellar mass $M_\ast$ for central galaxies in the local universe (red) and at $z=1$ (blue). The black solid line traces the evolutionary track of the galaxy, with circles color-coded by sSFR. Four specific redshift snapshots—$z=0, 0.2, 0.5$, and $1$—are indicated by arrows and labels.}
    \label{fig:color_single}
\end{figure}

Galaxy color strongly correlates with the star formation activity. As the SFR rises and falls during the episodic evolution, the color also changes accordingly. Fig.\,\ref{fig:color_single} shows the evolutionary history of an example galaxy (ID-433317) in the color-mass diagram. The two contours in this figure indicate the distribution of the $g-r$ color as a function of stellar mass $M_\ast$ for all central galaxies at $z=0$ (red contour) and at $z=1$ (blue contour). The $g-r$ color is calculated from the ratio of the total g- and r-band luminosities based on all stellar particles within the galaxy. In each case, four contour lines indicate the regions that include 10\%,  30\%, 60\%, and 90\% of the total sample. 
The solid line indicates the evolutionary track of the galaxy, color-coded by sSFR. The track shows an overall reddening trend, accompanied by multiple rapid variations, particularly since $z = 1$. These rapid changes are closely linked to variations in sSFR, exhibiting a clear anticorrelation.

It is worth noting that the $g-r$ color of this galaxy reaches its global peak at $z\approx0.2$, when its SFR was 0.46 dex below the main sequence at that time. Compared to the overall color distribution at that time, this galaxy used to be much redder than its star-forming counterparts and could even be identified as a `green valley' galaxy \citep{GreenValley_Salim2014, GreenValley_Mendez2011ApJ, Smith2022MNRAS}. However, soon after that, it rejuvenated and returned to the blue cloud once its star formation was ignited again. Among our present-day star-forming sample of 3644 galaxies, 47\% have had their historical SFR 0.5 dex below the main-sequence ridge (at the corresponding redshift) at least once since $z= 1$, and rejuvenated later to get back to the SFMS, or even above the ridge, at $z=0$.

\subsection{Temporal variation in the cold gas content}\label{sec:result_coldgas}

\begin{figure}[t!]
    \centering
    \includegraphics[width=\linewidth]{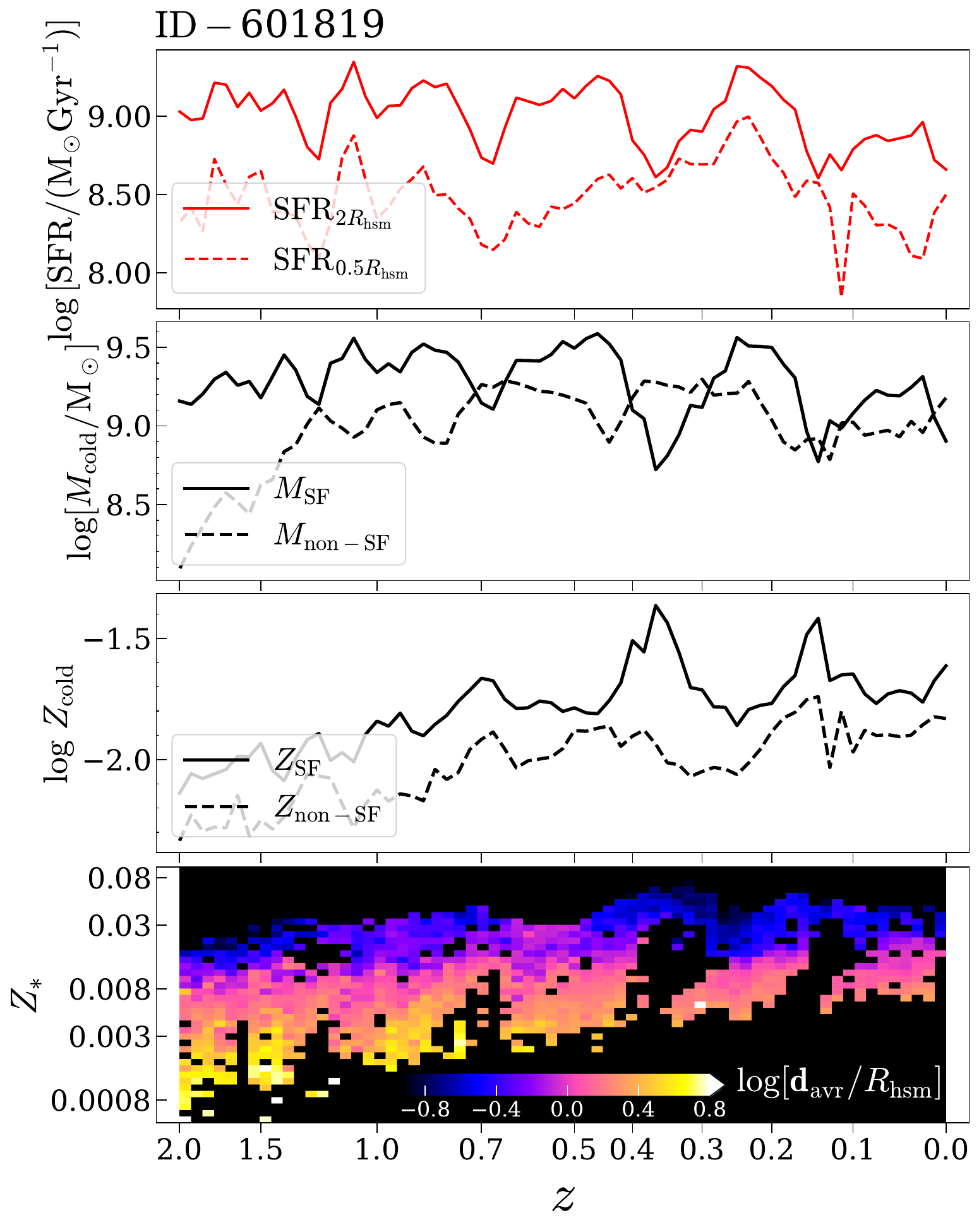}
    \caption{The episodic SFH and property evolution for one sample galaxy (ID-601819). The top panel presents the temporal variation in ${\rm SFR}_{\leqslant 2R_{\mathrm{hsm}}}$ (global, in solid line) and ${\rm SFR}_{\leqslant 0.5R_{\mathrm{hsm}}}$ (central, in dashed line). The second panel presents the evolution of cold gas mass within $2R_{\mathrm{hsm}}$ for the {\it star-forming} gas ($M_{\mathrm{SF}}$, solid line) and the cold {\it non-SF} gas ($M_{\mathrm{cold,\,non-SF}}$, dashed line). The third panel shows the metallicity evolution for both {\it star-forming} gas and cold {\it non-SF} gas, denoted as $Z_{\mathrm{SF}}$ (solid) and $Z_{\mathrm{cold,\,non-SF}}$ (dashed). The bottom panel presents the distribution of newly formed stars at redshift $z$ with stellar metallicity $Z_{\ast}$. The color represents the mass-weighted average distance (from the galaxy center, normalized by $R_{\mathrm{hsm}}$). We only plot those pixels that contain at least three stellar particles to avoid shot noise. }
    \label{fig:Mgas_Zgas_single}
\end{figure}

\begin{figure*}[ht!]
    \flushleft
    \includegraphics[width=0.98\textwidth]{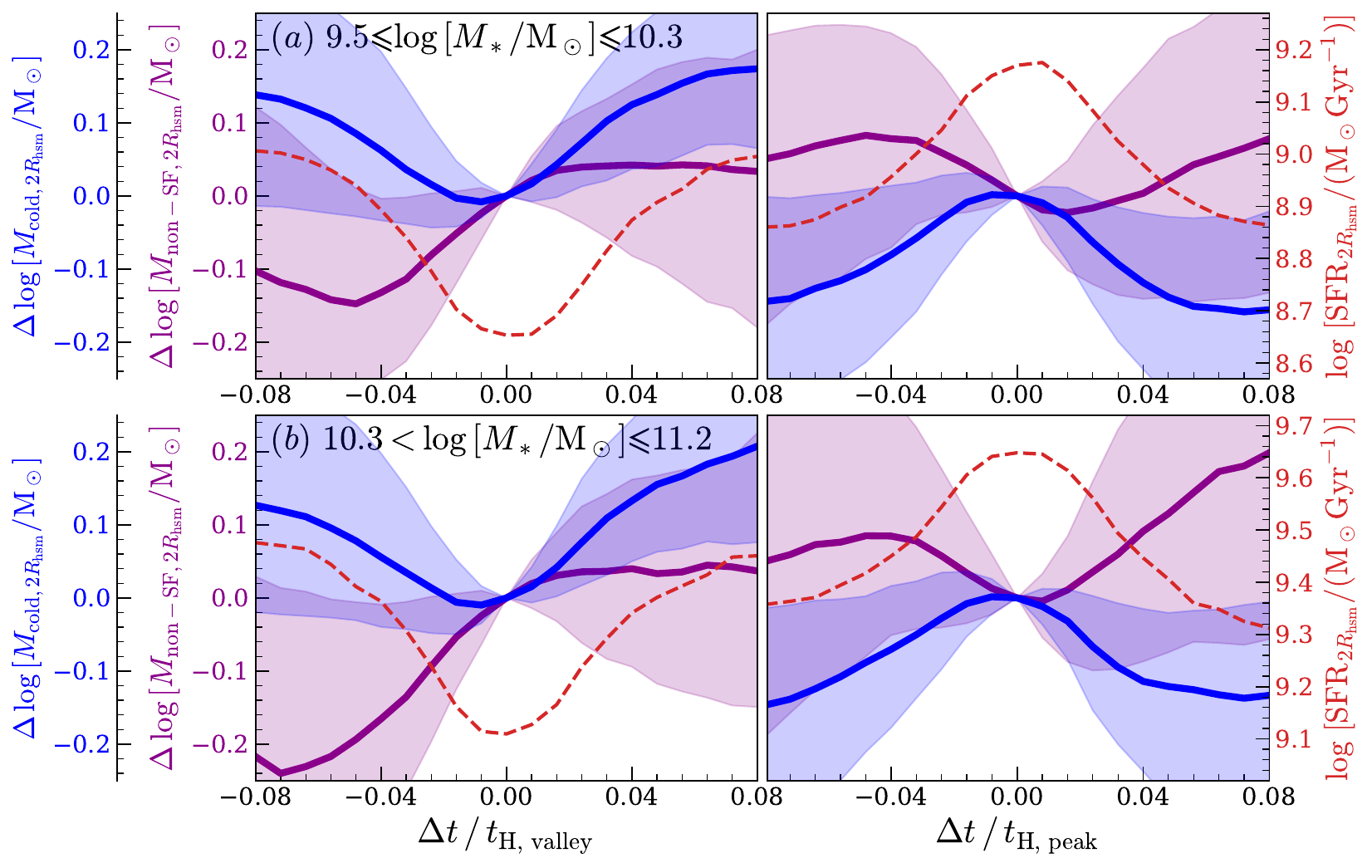}
    \caption{{\it Upper panel:} the relative mass $\Delta \log M$ of both total cold gas (blue) and cold but {\it non-SF} gas (magenta) within $2R_{\rm hsm}$ as a function of normalized time offset around the nearest peak/valley moment for lower-mass galaxies with mass $9.5 \leqslant \log M_{\ast,2R_{\rm hsm}}/\mathrm{M_\odot} \leqslant 10.3$. The time normalization procedure is described in the main text. In both panels, the shaded regions represent the central 68th percentile distribution. For clarity of visualization, the red dashed line marks the median SFR within $2R_{\mathrm{hsm}}$. {\it Lower panel:} same as the upper panel, but for higher-mass galaxies with mass $10.3 < \log M_{\ast,2R_{\rm hsm}}/\mathrm{M_\odot} \leqslant 11.2$.}
   \label{fig:gasZ_nearpeva_seperated}
\end{figure*}

As was investigated in \cite{SenWang_2022}, the rise and fall of SFR react (with a time delay) to the change in the circumgalactic cold gas reservoir going out to $\sim$100 kpc. To get an overview of this episodic behavior, we present the time evolution of the cold gas content and gas-phase metallicity along with SFR since $z\sim 2$ for an example galaxy (ID-601819) in Fig.\,\ref{fig:Mgas_Zgas_single}. The top panel presents the temporal variation in ${\rm SFR}_{\leqslant 2R_{\mathrm{hsm}}}$ (global, in solid line) and ${\rm SFR}_{\leqslant 0.5R_{\mathrm{hsm}}}$ (central, in dashed line). The second panel presents the evolution of cold gas mass within $2R_{\mathrm{hsm}}$ for the {\it star-forming} gas ($M_{\mathrm{SF}}$, solid line) and the cold {\it non-SF} gas ($M_{\mathrm{cold,\,non-SF}}$, dashed line). By definition, star-forming gas cells will transform some mass into stars at a certain rate. Therefore, the evolutions of ${\rm SFR}_{\leqslant 2R_{\mathrm{hsm}}}$ and $M_{\rm SF}$ are well in sync, which is clearly shown in the top two panels of Fig.\,\ref{fig:Mgas_Zgas_single}. However, the evolution of ${\rm SFR}_{\leqslant 2R_{\mathrm{hsm}}}$ is not fully synchronized with that of $M_{\mathrm{cold,\,non-SF}}$, but falls behind. As will be shown later, such time delays between the two cold gas budgets, as well as between the global SFR and the cold {\it non-SF} gas content, are universally present during the episodic star-formation evolution of galaxies -- a natural consequence of the cold {\it non-SF} gas reservoir taking time to gradually build up and form stars through compaction.

The third panel shows the evolution of metallicity in both {\it star-forming} gas and cold {\it non-SF} gas, denoted as $Z_{\mathrm{SF}}$ (solid) and $Z_{\mathrm{cold,\,non-SF}}$ (dashed), respectively. It is interesting to note that the three sets of major peaks of the gas metallicity at around $z=0.7$, $0.4$ and $0.2$ roughly correspond to the three valley features in $\mathrm{SFR}_{\leqslant 2R_{\mathrm{hsm}}}$, while the two wide valleys in gas metallicity (one between $z=0.7$ and $0.4$ and one between $z=0.4$ and $0.2$) also happen to correspond to the two major peaks in $\mathrm{SFR}_{\leqslant 2R_{\mathrm{hsm}}}$ in the top panel. If the evolutionary curves of the gas-phase metallicity are vertically flipped, they would largely resemble that of $\mathrm{SFR}_{\leqslant 2R_{\mathrm{hsm}}}$ (and in fact also the cold gas content in the second panel). This mirror behavior is especially notable for the metallicity of the {\it star-forming} gas (see also \citealt{Torrey2019}). This already indicates a close connection between gas metallicity, SFR, and the cold gas budget. We will investigate how such an episodic star-formation behavior can explain the observed fundamental metallicity relation \citep{Mannucci2010} in the upcoming paper (X. Mei et al. in preparation).

To further reveal statistical trends in our refined galaxy sample that are similar to those exhibited by the example galaxy in Fig.\,\ref{fig:Mgas_Zgas_single}, we stack the relative change of cold gas mass within $2R_{\mathrm{hsm}}$ — relative to its values at peaks and valleys — as a function of time offset around these critical moments. The results are shown in Fig.\,\ref{fig:gasZ_nearpeva_seperated}.
The stacking is performed separately for the two mass ranges defined in Section \ref{sec:method_peakvalley}. Since the evolution of {\it star-forming} gas mass closely traces that of SFR, we only plot relative change in the total cold gas, $\Delta\log M_{{\rm cold}}$ (blue), and that in the cold {\it non-SF} gas, $\Delta\log M_{{\rm non-SF}}$ (magenta). Due to the wide redshift range of the peaks and valleys, we normalize the time intervals by the Hubble time $t_{\rm H}$ at each peak/valley, so that time intervals at lower redshifts can be properly compared with those at higher redshifts. For comparison, we also present the results using absolute time (years) in Appendix \ref{sec:appendix_stacking}.

In both mass ranges, the rise and fall of the global SFR is roughly symmetric in time, i.e., the time from valley to peak is about the same as that from peak to valley. 
For the total cold gas content, it almost traces the SFR evolution, but precedes it by a small interval of $\sim 0.01 t_{\rm H}$.
In contrast, the rise and fall of the relative cold {\it non-SF} gas mass $\Delta\log M_{{\rm non-SF}}$ shows a significant time shift ahead of SFR. For the lower-mass galaxies,  the upturn (downturn) of $\Delta\log M_{\rm non-SF}$ from its minimum (maximum) starts $\sim 0.04-0.06\ t_{\rm H}$ before the upturn (downturn) of SFR. This time span equals $\sim 0.7$ Gyr at $z=0$ and $\sim 0.4$ Gyr at $z=1$. For the higher-mass galaxies, the corresponding timescales become $\sim 1$ Gyr at $z=0$ and $\sim 0.6$ Gyr at $z=1$\footnote{The exact value of such a time delay depends on the adopted threshold of $\Delta\log\,[\mathrm{SFR}/\mathrm{M_\odot Gyr^{-1}}]$ (see Section \ref{sec:method_SelESFG} and Appendix \ref{sec:appendix_threshold} for details). We refer the reader to the next paper in this series (X. Mei et al. in preparation) for a more detailed time analysis.}. As can be seen, both the advance in time of $\Delta\log M_{\rm non-SF}$ to SFR and the level of $\Delta\log M_{{\rm non-SF}}$ in the case of more massive galaxies are markedly larger than for lower-mass galaxies, indicating more efficient gas accretion in deeper gravitational potentials. We note to the reader that in the following analyses, as the results between the two mass ranges are qualitatively the same, we present only the combined results below.

As can be seen from Fig.\,\ref{fig:gasZ_nearpeva_seperated}, the cold {\it non-SF} gas reservoir (purple) begins to build up even while the total cold gas mass (blue) and SFR (red dashed) continue to decline toward the valleys. This indicates the coexistence of the inflow of fresh cold gas ({\it non-SF}) and the depletion of the {\it star-forming} gas pool due to feedback. With time, as the {\it non-SF} gas reservoir accumulates and compacts, a significant fraction of the cold gas in this phase is pushed above the star-forming density threshold and converted into {\it star-forming} gas, further driving the rise in SFR after the valley moment. From the SFR valley to the subsequent peak, the fraction of {\it star-forming} gas relative to total cold gas increases, rising from a median of $\lesssim$ 50\% at the valley to $\gtrsim 80\%$ at the peak.

Shortly after the SFR begins rising from its valley, the relative change in the cold {\it non-SF} gas mass (purple) peaks and then starts to decline, even though the total cold gas mass (blue) continues to increase. This suggests that while an increasing amount of gas is cooling from the hot phase and joining (thereby building up) the total cold gas budget, more and more {\it non-SF} cold gas is simultaneously being compressed to higher densities to feed the direct {\it star-forming} gas pool, further boosting the SFR -- though some of this {\it non-SF} gas may also leave this reservoir, especially as feedback becomes more effective. Over time, the cumulative effect of feedback eventually reduces the total cold gas budget, directly causing the SFR to decline after its peak. During the period from the SFR peak to the subsequent valley, both star formation and feedback continue to weaken, allowing the {\it non-SF} cold gas reservoir to gradually build up again, thereby fueling and soon initiating another episode of star formation.

The evolution of $\Delta\log M_{\rm cold}$ and SFR around SFR peaks and valleys, as shown by the blue and red curves in Fig.\,\ref{fig:gasZ_nearpeva_seperated} also reveals important difference in depletion time $\tau \equiv M_{\rm cold} / {\rm SFR}$ (or star formation efficiency ${\rm SFE} \equiv {\rm SFR} / M_{\rm cold}$) at these two types of moments. Quantitatively, at SFR valleys, the average depletion time is $\log\,[\tau/{\rm Gyr}] \sim 0.5{\pm}0.3$, and the average total cold gas mass is $\log\,[M_{\rm cold}/{\rm M}_{\odot}] \sim 9.4\pm 0.3$. At SFR peaks, these values are $\log\,[\tau/{\rm Gyr}] \sim 0.2{\pm}0.2$ and $\log\,[M_{\rm cold}/{\rm M}_{\odot}] \sim 9.7\pm 0.3$, respectively. Both quantities differ between the peak and valley moments by more than $1\sigma$. The rate of change in SFR does not exactly follow that of $M_{\rm cold}$. As can be seen, around SFR valleys, SFR drops faster and more substantially than $M_{\rm cold}$, implying a longer depletion time (or lower SFE) toward the valley moment. Conversely, around SFR peaks, SFR rises faster and more substantially than $M_{\rm cold}$, indicating a shorter depletion time (or higher SFE) toward the peak moment. This demonstrates that global changes in the total cold gas budget alone do not fully explain the rise and fall of SFR. At peak and valley moments, different microscopic mechanisms that operate at different levels must subtly modulate the efficiency of gas compaction for star formation.

\subsection{Two branches of star formation and the inward retreat of star formation in each star-forming episode}\label{sec:result_2branches}

As galaxies go through an episodic SFH, their sizes and metallicity distributions associated with the newly formed stars also evolve accordingly. In the bottom panel of Fig.\,\ref{fig:Mgas_Zgas_single} we present the radial distribution of newly formed stars at redshift $z$ with stellar metallicity $Z_{\ast}$ (for the example galaxy ID-601819). The color represents the mass-weighted average distance (from the galaxy center, normalized by $R_{\mathrm{hsm}}$). For this 2D histogram, we only plot those pixels that contain at least three stellar particles to avoid shot noise. Several episodes of star formation are clearly present.

For this example galaxy, in each star formation cycle (defined as between two adjacent SFR valleys that sandwich a SFR peak at least 0.5 dex higher than the valleys, see Section \ref{sec:method_SelESFG} for details) since $z\sim 1$, two branches of star formation exist. One branch is related to central star formation ($\la 0.5$ kpc) in higher-metallicity gas, which can always be on. In the second branch, a progression in a retreating/shrinking fashion is clearly present: a newer generation of stars very often forms initially at galaxy outskirts (at a few tens of kiloparsecs) as a result of cold gas accretion, which first arrives in disks at larger radii. 
With time, as the SFR gradually goes through the peak and starts decreasing, this round of star formation progressively born out of ever more metal-enriched gas, retreats inward to the inner $\sim 1$ kpc region. Finally, toward the end of this episode, the next round of cold fresh gas accretion already kicks off at the galaxy's outskirts, where new stars are born with lower metallicity. As a result, the intermittent pattern in the SFR is in fact more evident at galaxy outskirts (as indicated by yellow shading), while star formation in the very central regions of galaxies (indicated by blue shading) does not necessarily cease even at valleys of the global SFR. This can also be seen from the top panel of Fig.\,\ref{fig:Mgas_Zgas_single}, where the episodic evolution of ${\rm SFR}_{\leqslant 2R_{\mathrm{hsm}}}$ is much more significant than that of ${\rm SFR}_{\leqslant 0.5R_{\mathrm{hsm}}}$.

\begin{figure}[t!]
    \centering
    \includegraphics[width=\linewidth]{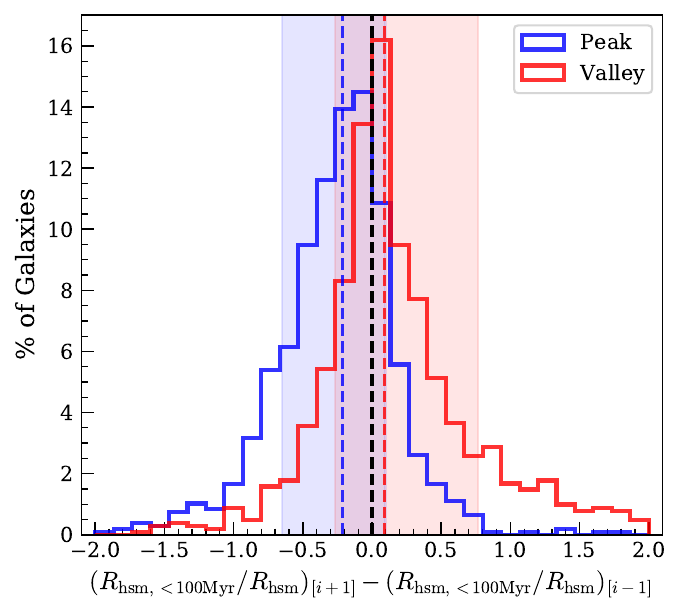}
    \caption{Histograms of the increment in normalized stellar half-mass radius of young stars $R_{\mathrm{hsm,\,<100\,Myr}}/R_{\mathrm{hsm}}$ from the $(i-1)^{\rm th}$ snapshot to the $(i+1)^{\rm th}$ snapshot, where $i$ is a snapshot at either SFR valley or peak. The blue and red histograms represent the increment in normalized size around peaks and valleys, respectively. The dashed lines and shaded regions represent the corresponding median and central 68th percentile in each histogram. The black dashed line indicates the value of zero (representing no change in $R_{\mathrm{hsm,\,<100\,Myr}}/R_{\mathrm{hsm}}$). Typically, the disk size of young stars grows around the time of the valley and becomes smaller around the time of the peak.}
    \label{fig:hsmr_concentration_peva}
\end{figure}

\begin{figure*}[ht!]
    \centering
    \includegraphics[width=0.48\linewidth]{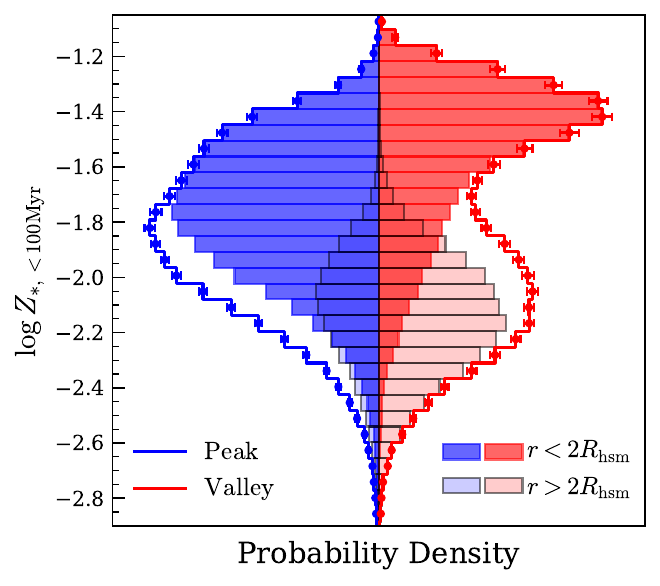}
    \includegraphics[width=0.48\linewidth]{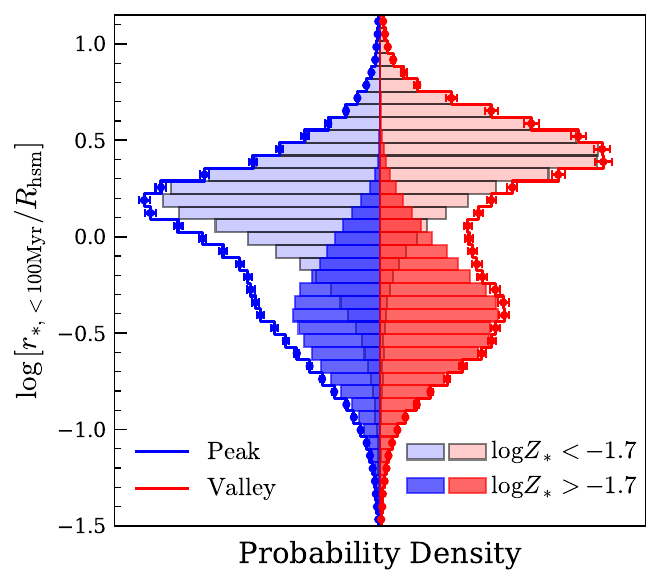}
    \caption{{\it Left panel:} the mass-weighted metallicity distribution of young stars, $\log\, Z_{\ast,\,\mathrm{<100Myr}}$, stacked at SFR peaks (blue) and valleys (red). The error bars represent the standard errors of the mean. The darker and lighter shaded regions represent the distributions of average young stellar metallicity calculated within and outside $2R_{\mathrm{hsm}}$. {\it Right panel:} the mass-weighted distribution of normalized galactocentric radii of young stellar, $\log\, [r_{\ast,\,\mathrm{<100Myr}}/R_{\mathrm{hsm}}]$. The darker and lighter shaded regions represent the distributions calculated with $\log\,Z_{\ast,\,\mathrm{<100Myr}}>-1.7$ and $<-1.7$, respectively.}
    \label{fig:youngstellarZ_hist_peva}
\end{figure*}

\begin{figure*}[ht!]
    \centering
    \includegraphics[width=\textwidth]{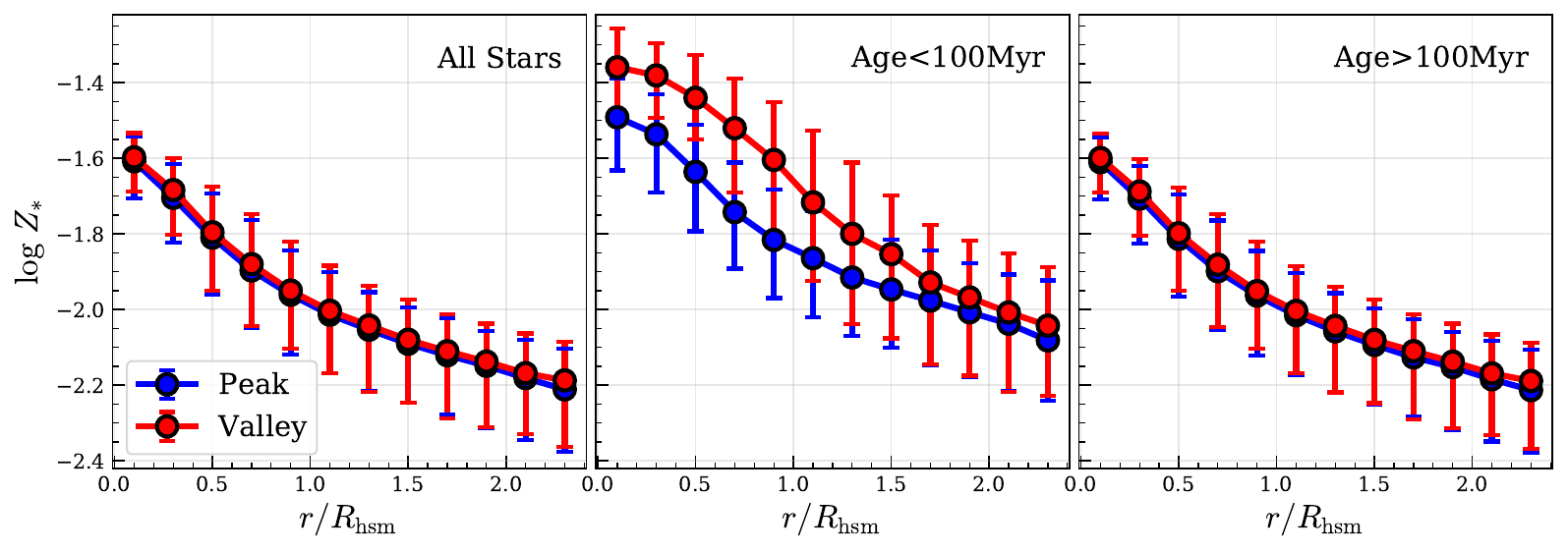}
    \caption{The radial profiles of the stellar metallicity from all stars (left), stars with ages younger than 100 Myr (middle), and stars older than 100 Myr (right). In each panel, the blue and red colors correspond to the time of peak and valley, respectively, where the error bars show the central 68th percentile distributions. For the younger stellar population, the metallicity profiles are less concentrated at SFR peaks compared to valleys.}
    \label{fig:stellarZ_profile_peva}
\end{figure*}

Such a two-branch star-formation pattern, as well as a retreating/shrinking fashion of the outer star-formation branch during each episode, as demonstrated by the example galaxy, have been broadly observed among the progenitors of all present-day star-forming galaxies (regardless of the adopted $\Delta\log\,[\mathrm{SFR}/\mathrm{M_\odot Gyr^{-1}}]$ threshold for the refined galaxy sample). Statistically, we shall also expect to see that the size of the stellar disk as measured by young stars shall grow around SFR valleys due to the arrival of fresh gas supply first at larger distances, and shrink around SFR peaks as star formation retreats to the central region. This is clearly demonstrated using galaxy sizes at the identified peaks and valleys for the refined galaxy sample in Fig.\,\ref{fig:hsmr_concentration_peva}, which presents the histograms of the relative size increment $\Delta [R_{\mathrm{hsm,\,<100\, Myr}}/R_{\mathrm{hsm}}]$ of stellar disks from the $(i-1)^{\rm th}$ snapshot to the $(i+1)^{\rm th}$ snapshot, where $i$ is a snapshot number at either SFR valleys or peaks. Here, we take all stellar particles younger than 100 Myr by that time to calculate their half-mass radius $R_{\mathrm{hsm,\,<100\, Myr}}$ and normalize this size by $R_{\mathrm{hsm}}$ to eliminate systematic differences between galaxies. As can be seen, on average, galaxies (as traced by newly formed stars) tend to grow larger around SFR valleys and become smaller around SFR peaks. We note that using $R_{200}$ (the radius of the galaxy halo within which the average total matter density is 200 times the critical density of the universe at that time) for normalization has no qualitative impact on the result. We mention in passing that the distribution of newly formed/young stars may thus vary during different epochs of star formation. Such a temporal variation may potentially contribute to scatter in galaxy sizes, particularly as observed at UV wavelengths.

\begin{figure*}[ht!]
    \centering
    \includegraphics[width=\textwidth]{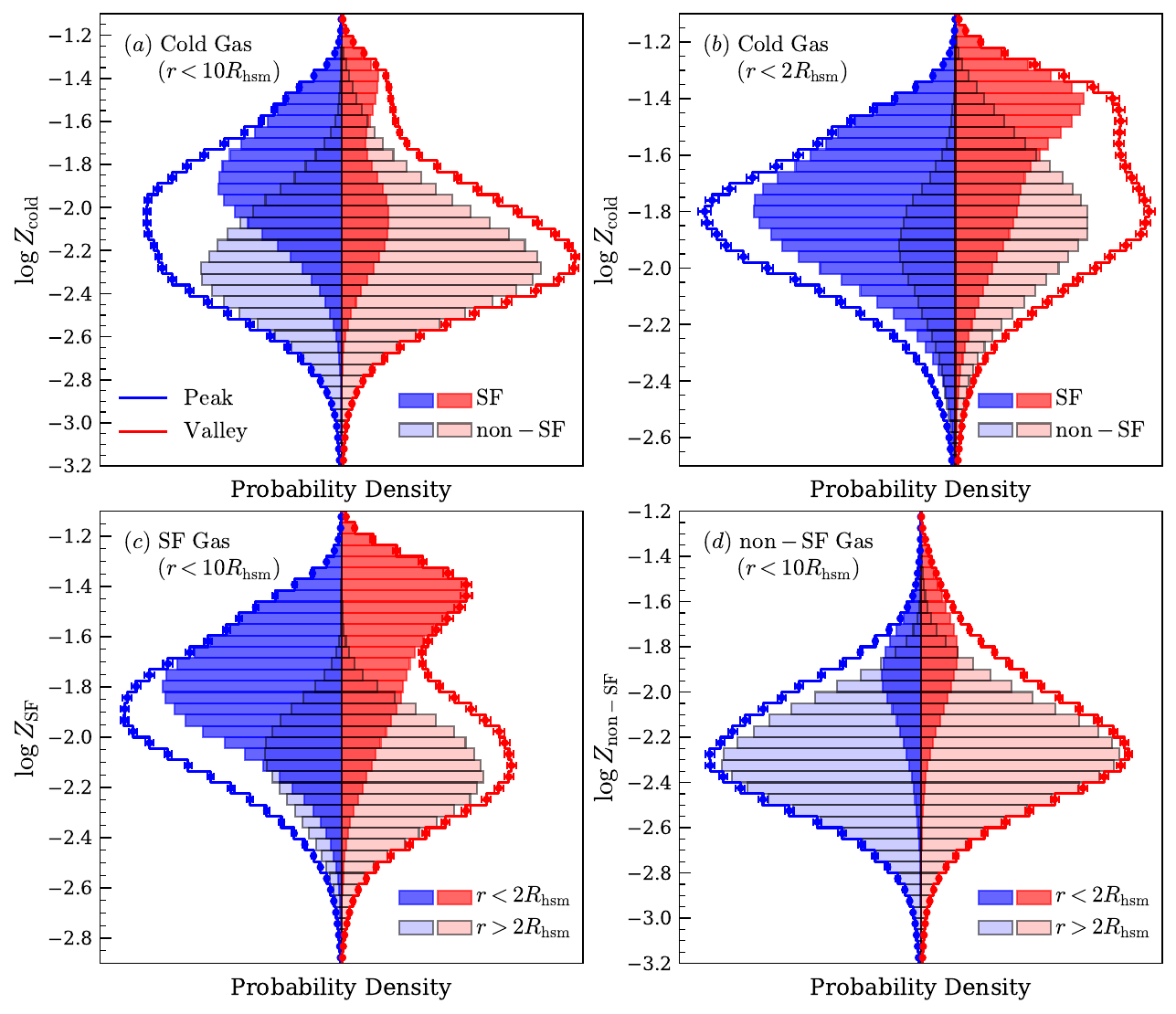}
    \caption{{\bf (a)} Stacked histogram of average mass-weighted total cold gas metallicity, $\log Z_{\rm cold}$, at SFR peaks (blue) and valleys (red) within $10R_{\mathrm{hsm}}$. The darker and lighter shaded regions represent the distributions calculated for {\it star-forming} and cold {\it non-SF} gas, respectively. Plotting style similar to Fig.\,\ref{fig:youngstellarZ_hist_peva}. {\bf (b)} Similar to panel (a), but evaluating within $2R_{\mathrm{hsm}}$. {\bf (c)} Stacked histogram of average mass-weighted {\it star-forming} gas metallicity, $\log Z_{\rm SF}$, at SFR peaks (blue) and valleys (red) within $10R_{\mathrm{hsm}}$. The darker and lighter shaded regions represent the metallicity histograms within and outside $R=2R_{\mathrm{hsm}}$, respectively. {\bf (d)} Similar to panel (c), but calculated for metallicity $\log Z_{\rm non-SF}$ of the cold {\it non-SF} gas.}
    \label{fig:coldgasZ_hist_peva}
\end{figure*}

\subsection{Differential metallicity distributions of gas and young stars between SFR peaks and valleys}
\label{sec:result_diffmet}

When further taking metallicity into account, the two branches of star formation during each episode are clearly revealed by the metallicity distributions of the younger stellar population and of the cold gas. The former is presented in Fig.\,\ref{fig:youngstellarZ_hist_peva}. Here again, `young stars' are approximated by stellar particles with age younger than 100 Myr, and their metallicity is denoted as $\log Z_{\ast,\,<100\,{\rm Myr}}$. In the left panel, the blue and red histograms present the distributions of $\log Z_{\ast,\,<100\,{\rm Myr}}$ at SFR peaks and valleys, respectively. The darker and lighter shaded areas represent the distributions calculated within and outside $2R_{\mathrm{hsm}}$, respectively. As can be seen, the overall distribution of $\log Z_{\ast,\,<100\,{\rm Myr}}$ measured at SFR valleys clearly exhibits bimodality, with a higher (lower) metallicity peak corresponding to the inner (outer) star-formation region. To the contrary, the overall distribution of $\log Z_{\ast,\,<100\,{\rm Myr}}$ at SFR peaks exhibits a broad shape which is largely dominated by star formation within $2R_{\mathrm{hsm}}$. In the right panel, the histograms present the distributions of galactocentric distance $\log\,[r_{\ast,\,<100\,{\rm Myr}}/R_{\mathrm{hsm}}]$ of young stars at SFR peaks and valleys. The darker and lighter shaded areas now represent the radial distributions of young stars with $\log Z_{\ast,\,<100\,{\rm Myr}}>-1.7$ and $<-1.7$, respectively. As can be seen, at both epochs (peak and valley), the distributions exhibit bimodality to a certain degree. Young stars with lower metallicity tend to reside in galaxy outskirts (peak at $2-3R_{\mathrm{hsm}}$), while young stars with higher metallicity tend to occupy the central regions (peak at $\sim 0.3R_{\mathrm{hsm}}$).

Such behaviors of the young star metallicity distribution originate from their different spatial distributions at SFR peaks and valleys due to the presence of two star-formation branches. It is worth noting that the younger stellar population at SFR peaks is, in general, less concentrated and of lower metallicity (in particular at distances within $2R_{\mathrm{hsm}}$), in comparison to SFR valleys. This can be more clearly seen in Fig.\,\ref{fig:stellarZ_profile_peva}, which presents the radial profiles of stellar metallicity from all stars (left), stars younger than 100 Myr (middle), and stars older than 100 Myr (right). Blue and red colors correspond to SFR peaks and valleys, respectively. As can be seen, metallicity profiles from the entire stellar population are dominated by the older stellar population and show no difference between the peak and valley moments. However, for the younger stellar population, a clear difference is observed: within $2R_{\mathrm{hsm}}$, young star metallicity is significantly lower at SFR peaks than at valleys -- a natural consequence of stars born out of the freshly replenished gas with lower metallicity.

The gas-phase metallicity distributions presented in Fig.\,\ref{fig:coldgasZ_hist_peva} also reveal the presence of two star-formation branches during the episodic evolution. The top panels show the stacked mass-weighted total cold gas metallicity, $\log Z_{\rm cold}$, at SFR peaks (blue) and valleys (red) within $10R_{\mathrm{hsm}}$ and $2R_{\mathrm{hsm}}$, respectively. dThe darker and lighter shaded regions represent the distributions calculated for {\it star-forming} and cold {\it non-SF} gas, respectively. As can be seen, within a radius of $10\,R_{\mathrm{hsm}}$ from the galaxy center, the cold but {\it non-SF} gas contributes a significantly higher fraction to the total cold gas budget than the {\it star-forming} gas does -- particularly at SFR valleys, where it dominates. Specifically, {\it non-SF} gas accounts for $51^{+20}_{-19}$\% and $77^{+12}_{-16}$\% of the total cold gas within $10\,R_{\mathrm{hsm}}$ at SFR peaks and valleys, respectively. Within $2\,R_{\mathrm{hsm}}$, {\it star-forming} gas dominates the total cold gas budget at SFR peaks, while the cold {\it non-SF} gas still contributes a substantial fraction at SFR valleys (which can also be seen in Fig.\,\ref{fig:Mgas_Zgas_single}). Specifically, {\it non-SF} gas accounts for $16^{+21}_{-10}$\% and $54^{+21}_{-30}$\% of the total cold gas within $2\, R_{\rm hsm}$ at SFR peaks and valleys, respectively. We note again that, in particular, at the valley moments, the cold {\it non-SF} gas reservoir has already started building up, prior to an enhanced star formation kicking off (see Fig.\,\ref{fig:gasZ_nearpeva_seperated}). In addition, the cold but {\it non-SF} gas in general exhibits a lower metallicity than the {\it star-forming} gas (although less so when evaluated within $2R_{\mathrm{hsm}}$ at SFR peaks), which is a result of the combination of fresh gas replenishment first at the cold {\it non-SF} gas reservoir as well as feedback and chemical enrichment to the {\it star-forming} gas pool.

The bottom two panels in the same plot show histograms similar to the top two panels, but evaluated for the {\it star-forming} (left) and the cold {\it non-SF} (right) gas within $10R_{\mathrm{hsm}}$, respectively. Here, the darker and lighter shaded regions represent the metallicity histograms within and outside $2R_{\mathrm{hsm}}$, respectively. As can be seen, for both gas components, gas located at galaxy outskirts ($r>2R_{\mathrm{hsm}}$) in general possesses lower metallicity than in the inner region ($r<2R_{\mathrm{hsm}}$), which is a behavior similar to taht of the younger stellar population. In addition, similar to the left panel of Fig.\,\ref{fig:youngstellarZ_hist_peva}, the metallicity of {\it star-forming} gas also exhibits a clear bimodal distribution at SFR valleys, composed of a higher-metallicity branch in the inner regions and a lower-metallicity branch at the outskirts. This is also the consequence of two branches of star formation, as already described above. It is also worth noting that within $2R_{\mathrm{hsm}}$, the metallicity distribution of {\it star-forming} gas biases toward higher values at SFR valleys (as it is dominated by the central star-forming branch) than at SFR peaks. This is, in fact, consistent with what one would expect from the fundamental metallicity relation (see X. Mei et al. in preparation). Finally, we also note to the reader that measuring the cold gas metallicity probability distribution over a wide radial range across a galaxy may provide good evidence of the star-formation phase that a galaxy is in, further evidencing an episodic star formation scenario.

\section{Results: Scatter of SFMS Galaxies}\label{sec:results-scatter}

\begin{figure}[t!]
    \centering
    \includegraphics[width=1\linewidth]{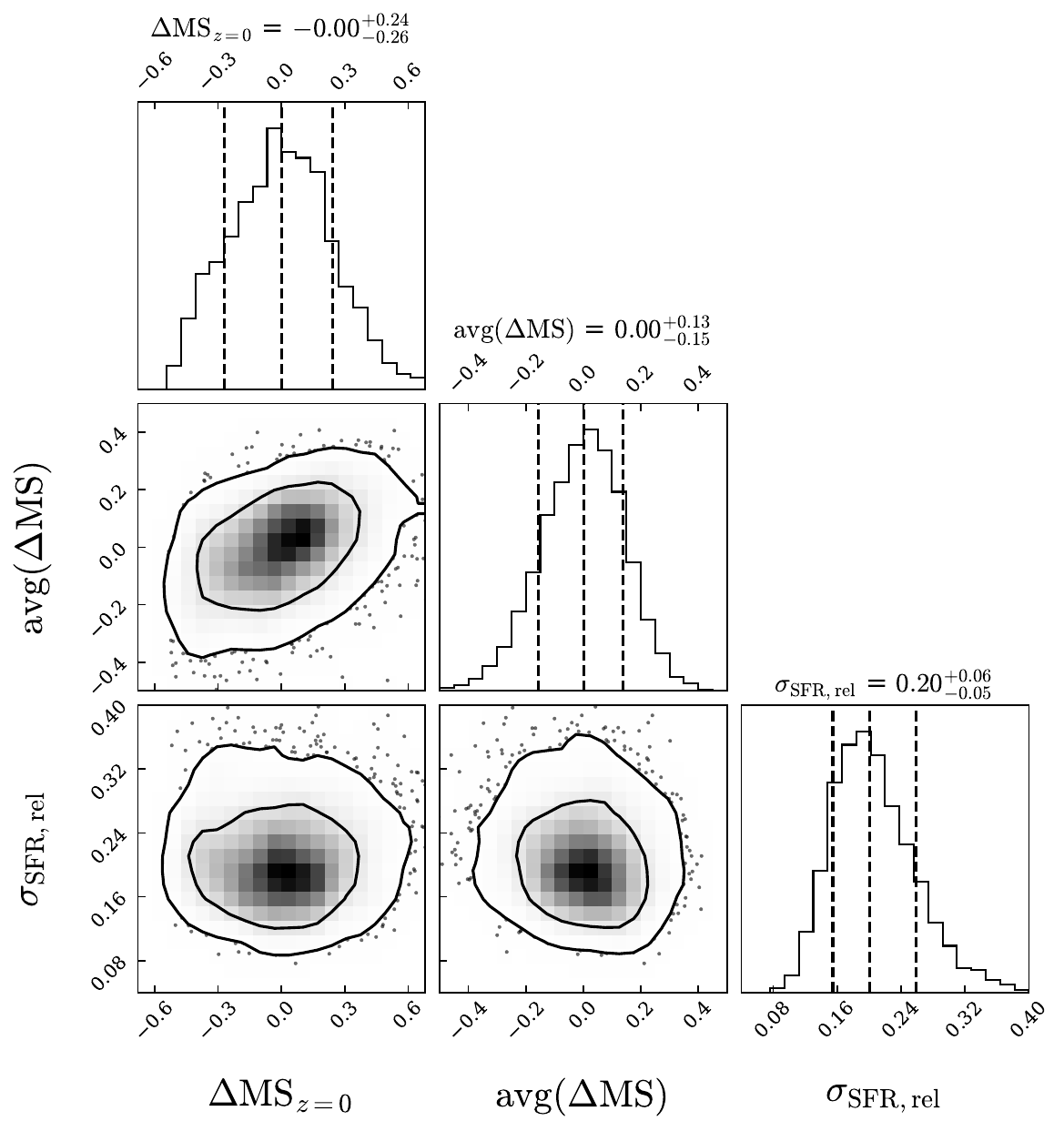}
    \caption{The correlation between the main-sequence offset at $z=0$ ($\Delta\rm MS_{z=0}$), the historically averaged main-sequence offset since $z=1$ ($\text{avg}(\Delta\rm MS)$) and the temporal fluctuation of individual galaxies since $z=1$ ($\sigma_{\rm SFR, rel}$). The SFR is measured within $2R_{\rm hsm}$. The 16th, 50th, and 84th percentiles of each quantity are shown above the corresponding panel.}
    \label{fig:cornerplot_1}
\end{figure}

\begin{figure*}[ht!]
    \centering
    \includegraphics[width=0.49\textwidth]{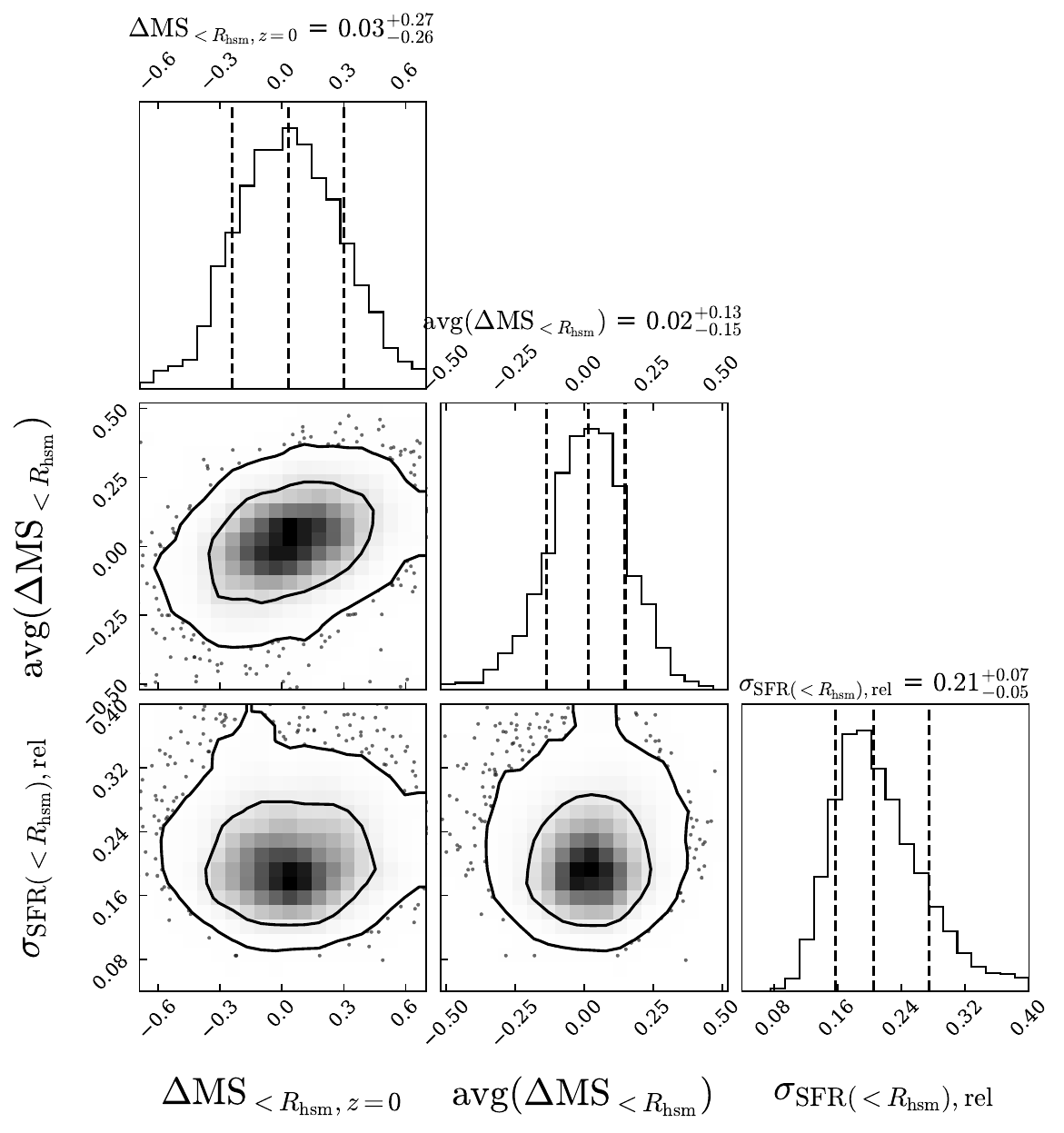}
    \includegraphics[width=0.49\textwidth]{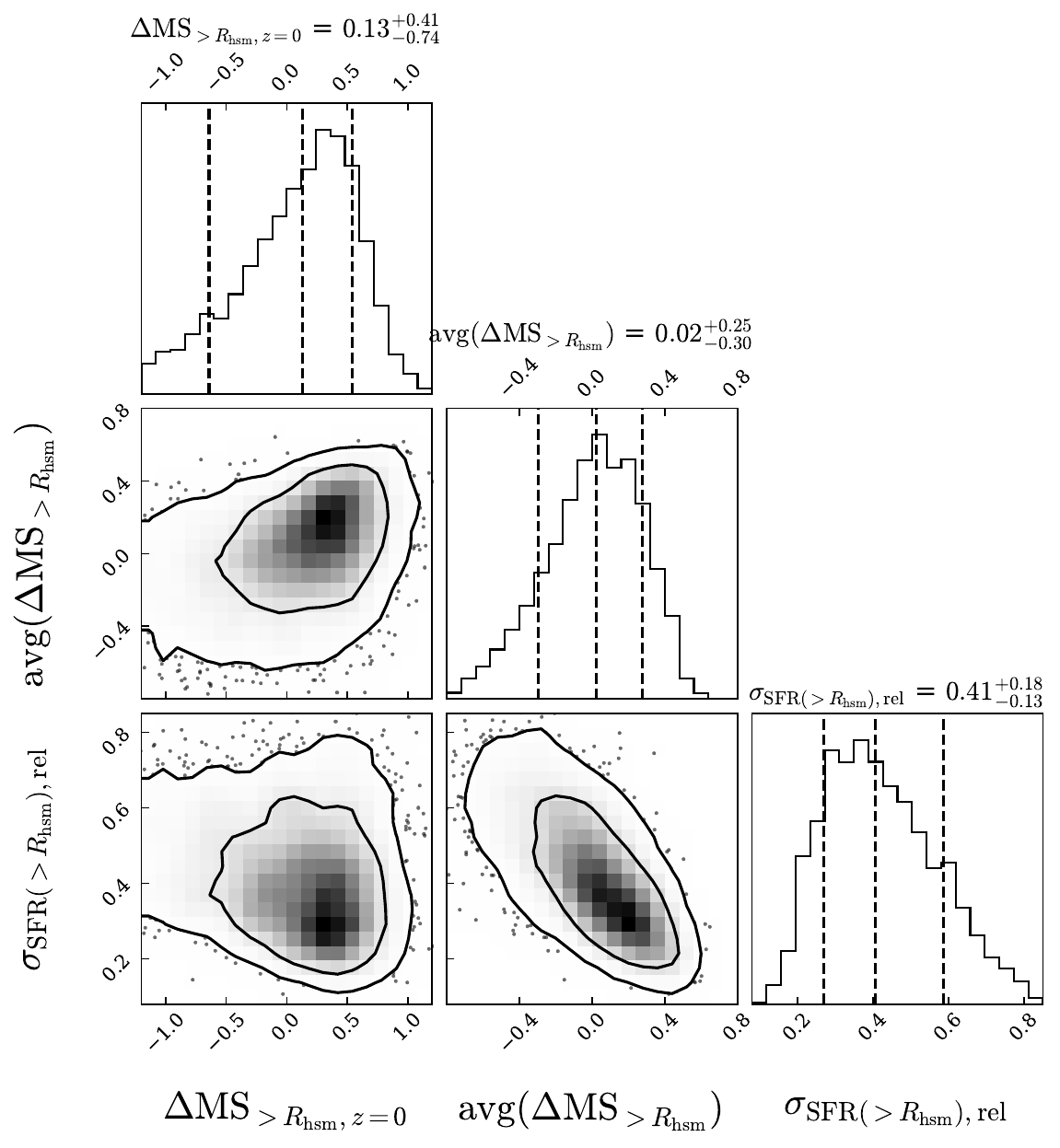}
    \caption{{\it Left panel:} similar to Fig.\,\ref{fig:cornerplot_1} except that the SFR is measured within $R_{\rm hsm}$. The main-sequence offset $\Delta\rm MS_{<R_{\rm hsm}}$ is recalculated from the newly fitted ridge, and the temporal SFR fluctuation of individual galaxies $\sigma_{\mathrm{SFR}(<R_{\rm hsm}), \mathrm{rel}}$ is also taken from the refitted long-term evolution for each galaxy. {\it Right panel:} similar to the left panel except that the SFR is measured in an annulus between $R_{\rm hsm}$ and $2R_{\rm hsm}$.}
    \label{fig:cornerplot_2}
\end{figure*}

Across the episodic SFH of present-day SFMS galaxies, many properties vary as SFR changes in time. However, these properties fluctuate more or less about a quasi-equilibrium state (e.g., \citealt{Forbes2014, Torrey2019, DeLucia2020, vanLoon2021}. Therefore, we expect that the episodic star-formation behavior causing temporal fluctuation in SFR should be able to (at least partially) explain the SFMS scatter today. For present-day ($z=0$) star-forming galaxies in the TNG100, we determine the ridge of the SFMS and report the 16th, 50th, and 84th percentiles of their main-sequence offset $\Delta \rm MS_{z=0}$ is $0.00^{\bf +0.24}_{\bf -0.26}$ dex.
In comparison, the 16th, 50th, and 84th percentiles of $\Delta\rm MS$ among all galaxy progenitors since $z=1$ is $0.01^{\bf +0.25}_{\bf -0.28}$ dex, exhibiting a scatter consistent with that of the SFMS at $z=0$. This likeness indicates that the SFMS scatter does not show appreciable evolution since then. 
As the evaluation of the intrinsic differentiation between galaxies, the 16th, 50th, and 84th percentiles of the historically averaged main-sequence offset since $z=1$, ${\rm avg}(\Delta\rm MS)$, is $0.00^{\bf +0.13}_{\bf -0.15}$ dex. 
In addition, the 16th, 50th, and 84th percentiles of the temporal SFR fluctuations within individual galaxies around their long-term average, $\sigma_{\rm SFR, rel}$, is 
${\bf 0.20}^{+0.06}_{-0.05}$ dex (for sample selection and detailed definitions, see Section \ref{sec:method_galaxysample}).
These statistics reveal that the SFMS scatter (on the order of 0.25 dex) can be well accounted for by the temporal fluctuation in SFR (on the order of 0.2 dex) within individual galaxies, plus a galaxy-to-galaxy differentiation (on the order of 0.15 dex) in long-term averaged SFR (see also \citealt{Matthee_and_Schaye2019, NIHAO_Blank2021, JennyWan2025}). Assuming the ergodicity of the SFMS, \citet{EnciWang_and_Lilly2022b} also found a larger contribution from the temporal fluctuation in SFR to the scatter of SFMS, based on the observed SFMS scatter derived on two different timescales.

It is then interesting to consider whether galaxies that live above (below) the main-sequence ridge today have, on average, lived above (below) the ridge in their history, and whether a strong correlation exists between the historically averaged main-sequence offset $\mathrm{avg}(\Delta\rm MS)$ and the temporal fluctuation $\sigma_{\rm SFR, rel}$. To answer this question, we present Fig.\,\ref{fig:cornerplot_1}. This corner plot shows the correlation between $\Delta\rm MS_{z=0}$, $\text{avg}(\Delta\rm MS)$ and $\sigma_{\rm SFR, rel}$. We find a positive correlation between present-day main-sequence offset $\Delta\rm MS_{z=0}$ and historical averaged main-sequence offset since $z \sim 1$ ($\mathrm{avg}(\Delta\rm MS)$), indicating that galaxies with higher (lower) $\Delta\rm MS$ at $z=0$ tend to reside above (below) SFMS from $z\sim 1$. This result echoes a recent observational result from \cite{JennyWan2025}, conducted for 1928 star-forming galaxies from the LEGA-C survey. In comparison, correlation between $\mathrm{avg}(\Delta\rm MS)$ and $\sigma_{\rm SFR, rel}$ is insignificant. We note that in this study, we aim at understanding the key processes during the episodic evolution and how this causes a given population of galaxies to appear with diverse observed properties. We have not answered what physical origins have led to the intrinsic differences among galaxies or what physical processes have largely determined the amplitude of the temporal fluctuation. We leave these investigations to future work of this series (Y. Gui et al. in preparation).

As presented in Section \ref{sec:result_2branches}, star-formation that happens at inner regions of galaxies and that at the outskirts also follow markedly different evolutionary patterns. A related question worth asking is then how the above-investigated SFR properties depend on the (projected) spatial region within which they are evaluated. To answer this question, we carry out the same analysis but considering the star formation within an aperture, ${\rm SFR}_{\rm apt}$, evaluated (1) within a radius of $R_{\rm hsm}$ from the galaxy center, and (2) within an annulus between $R_{\rm hsm}$ and $2R_{\rm hsm}$. In each case, the SFMS galaxy sample remains the same. However, we refit the main-sequence ridge at different redshifts using ${\rm SFR}_{\rm apt}$ (instead of global SFR), and re-calculate the offset from the ridge as ${\rm avg}(\Delta\rm MS_{\rm apt})$. For the temporal fluctuation $\sigma({\rm SFR}_{\rm apt})$, we refit a linear relation using $\log{\rm SFR}_{\rm apt}-\log(1+z)$ and take the linear prediction as the mean to calculate $\sigma({\rm SFR}_{\rm apt})$. 
Fig.\,\ref{fig:cornerplot_2} presents the results similar to Fig.\,\ref{fig:cornerplot_1} but for SFR evaluated within two different regions. The left panel shows the distributions and correlations with SFR measured within $R_{\rm hsm}$, while the right panel shows those with SFR evaluated in the annulus between $R_{\rm hsm}$ and $2R_{\rm hsm}$.

As can be seen, the results for ${\rm SFR}_{\rm apt}$ measured within $R_{\rm hsm}$ very much resemble those using global SFR (as presented in Fig.\,\ref{fig:cornerplot_1}), indicating that the global feature is largely dominated by the condition within $R_{\rm hsm}$. However, the results for ${\rm SFR}_{\rm apt}$ measured between $R_{\rm hsm}$ and $2R_{\rm hsm}$ differ substantially from the former cases. Firstly, $\Delta \rm MS_{z=0}$ no longer exhibits a symmetric distribution around $\Delta \rm MS =0$ (the new ridge), but with a long tail bias toward lower values in $\Delta \rm MS$. Secondly, using SFR evaluated between $R_{\rm hsm}$ and $2R_{\rm hsm}$ (i.e., at galaxy outskirts), the scatter of ${\rm avg}(\Delta\rm MS)$ as well as $\sigma_{\rm SFR, rel}$ have all become significantly larger than those evaluated within the central $R_{\rm hsm}$. This can be attributed to the fact that the episodic evolution at galaxy outskirts is more prominent than in inner regions (as demonstrated in Section \ref{sec:result_2branches}) thus causing larger scatters. Finally, we also mention in passing that for SFR measured at such larger radii, ${\rm avg}(\Delta\rm MS)$ and $\sigma_{\rm SFR, rel}$ now exhibit a strong negative correlation, i.e., galaxies with lower ${\rm avg}(\Delta\rm MS)$ tend to have a larger temporal fluctuation of $\sigma_{\rm SFR, rel}$. This result is likely attributed to the steep decline of SFR in the outskirts at some specific epochs, which elevates $\sigma_{\rm SFR, rel}$ and suppresses ${\rm avg}(\Delta\rm MS)$ simultaneously. We defer the investigation of the detailed physical origin to an upcoming paper (Y. Gui et al. in preparation).

We have also explored the Hurst parameter ($H$) as defined in \cite{Kelson2014}, which provides a quantitative measure of the temporal correlation structure of SFR fluctuations. We apply the detrended fluctuation analysis method to the sSFR residuals of our total star-forming central galaxy sample over $z =1$ to $z=0$, and find that the majority of galaxies exhibit $H > 0.5$, indicating that SFR fluctuations are positively correlated over time and carry long-term memory, which supports the interpretation that SFR variability is not purely random but is influenced by underlying physical processes with characteristic timescales.

\section{Conclusions and Discussions}\label{sec:conclusions}

This paper is a follow-up study of \cite{SenWang_2022}, with a more focused and detailed investigation of the episodic SFH of present-day SFMS galaxies since $z \sim 1$. The goal of this study is to present an overview of the episodic star formation scenario and investigate how well the resulting temporal variation in SFR can account for the observed scatter of the star-forming main sequence at $z=0$. This is achieved through tracing back in time the history of star-forming galaxies at $z=0$ in the TNG100 simulation and investigating the coevolution of the size, metallicity, and radial distribution of young stars and cold gas, as the galaxies go through SFR peaks and valleys in the past $7-8$ Gyr. In particular, we split the cold gas into a {\it non-star-forming} gas reservoir and a {\it star-forming} gas pool (see Section \ref{sec:method_properties} for detailed definitions and the method), and investigate the different stellar and gas properties at typical SFR peaks and valleys (see Section \ref{sec:method_peakvalley}). Our main results about the episodic SFH and its impact on scatter in SFMS are listed below:

\begin{itemize}
    \item The evolution of total cold gas mass and SFR (within $2R_{\rm hsm}$) around SFR peaks and valleys reveals key differences in depletion time $\tau \equiv M_{\rm cold} / {\rm SFR}$ or star formation efficiency (${\rm SFE}=\tau^{-1}$). At SFR valleys, the average depletion time is $\log[\tau / {\rm Gyr}] \sim 0.5$ and the average cold gas mass is $\log [M_{\rm cold}/{\rm M}_{\odot}] \sim 9.4$; at SFR peaks, the corresponding values are $\log[\tau / {\rm Gyr}] \sim 0.2$ and $\log [M_{\rm cold}/{\rm M}_{\odot}] \sim 9.7$. Both quantities differ between SFR peaks and valleys by more than $1\sigma$. SFR changes more dramatically than cold gas mass: SFR drops faster toward valleys (longer $\tau$ or lower SFE) and rises faster toward peaks (shorter $\tau$ or higher SFE). This indicates that the global cold gas budget alone cannot explain SFR variations; instead, microscopic mechanisms modulating gas compaction and star formation efficiency play a crucial role (see Section \ref{sec:result_coldgas}).

    \item During each episode, two branches of star formation exist. While one star-formation branch happens in heavily polluted gas in the inner region (and in many cases remains always on), a secondary star-formation branch starts from lower-metallicity regions at galaxy outskirts where fresh gas first arrives, and gradually retreats to the centers of galaxies. This picture is consistent with the coplanar inflow model proposed by \citet{EnciWang_and_Lilly2022b, EnciWang_and_Lilly2023a, EnciWang_and_Lilly2023b} and \citet{CheqiuLyu2025}. As a consequence, the intermittent pattern in SFR is more prominent at galaxy outskirts, while star formation in inner regions does not necessarily cease even at valleys of global SFR (see Fig.\,\ref{fig:Mgas_Zgas_single} and Section \ref{sec:result_2branches}).

    \item The relative stellar size of the younger stellar population $R_{\mathrm{hsm},\,<100{\rm Myr}}/R_{\mathrm{hsm}}$ systematically increases around SFR valleys, and decreases around SFR peaks, as a result of stellar disk growth (for the outer-branch star-formation) first at larger radii at the beginning of each star-formation episode (i.e., SFR valley), followed up by a `shrinkage' after the peak time (see Fig.\,\ref{fig:hsmr_concentration_peva} and Section \ref{sec:result_2branches}). This temporal variation in the radial extent of young stars is anticipated to contribute to the scatter in galaxy sizes as observed at UV wavelengths (although the latter may be modulated by spatially varying attenuation; see \citealt{JunkaiZhang2023}).

    \item Statistically, the younger stellar populations in galaxies at SFR {\it valleys} exhibit a strong bimodal probability distribution function (PDF) in metallicity, which is composed of a higher and a lower-metallicity peak corresponding to the inner ($r< 2R_{\mathrm{hsm}}$) and the outer ($r> 2R_{\mathrm{hsm}}$) star-formation region, respectively (Fig.\,\ref{fig:youngstellarZ_hist_peva}). The younger stellar population at SFR peaks possesses systematically lower metallicity than at SFR valleys across the galaxy radial range within $2R_{\mathrm{hsm}}$ as a result of fresh gas replenishment. The total stellar population, however, does not reveal noticeable differences between the peaks and valleys (see Fig.\,\ref{fig:stellarZ_profile_peva} and Section \ref{sec:result_diffmet}).

    \item Similarly, the two branches of star formation, along with the retreating fashion of disk growth as the SFR rises from valley to peak, also leave different observational imprints on the cold gas metallicity distribution. In particular, bimodal PDFs are prominent at SFR {\it valleys} -- both in the metallicity of the total cold gas (the sum of the cold {\it non-SF} gas reservoir and the {\it star-forming} gas pool) within $2R_{\rm hsm}$ and in the metallicity of {\it star-forming} gas within $10R_{\rm hsm}$ (i.e., measured across both inner and outer galactic regions). These bimodalities arise from the coexistence of distinct chemical signatures associated with two star-forming branches (Fig.\,\ref{fig:coldgasZ_hist_peva} and Section \ref{sec:result_diffmet}).

    \item The median SFMS offset ($\Delta \rm MS$) at $z=0$ is $0.00^{\bf +0.24}_{\bf -0.26}$ dex, where the uncertainties correspond to the 16th and 84th percentiles, and the historical main-sequence offset $\Delta \rm MS$ among all of galaxy progenitors since $z=1$ is $0.01^{\bf +0.25}_{\bf -0.28}$ dex, indicating that the MS scatter does not show appreciable evolution since then. Among individual galaxies and their progenitors since $z=1$, the historically averaged $\Delta \rm MS$ is 
    $0.00^{\bf +0.13}_{\bf -0.15}$ dex, while the $1\sigma$ of temporal fluctuation within each galaxy is $\Delta\rm MS$ is ${\bf 0.20}^{+0.06}_{-0.05}$ dex. These statistics support the idea that the galaxies are going through episodic SFH about a quasi-equilibrium state; the temporal fluctuation in SFR (on the order of 0.2 dex) together with intrinsic differentiation between galaxies (on the order of 0.15 dex) can explain $z=0$ SFMS scatter (on the order of 0.25 dex) (see Fig.\,\ref{fig:cornerplot_1} and Section \ref{sec:results-scatter}). 
    
\end{itemize}

We note that the conclusions presented above are entirely based on the TNG100 simulation. As the episodic star-formation behavior is a combined consequence of both external gas replenishment and internal feedback processes, the details of such behaviors are surely subject to the adopted feedback models by the simulation. In addition, the results also indicate that such an episodic star-formation scenario can partially explain the observed scatter of the SFMS. It will be highly interesting to see whether different cosmological simulations of this kind shall provide consistent answers and thus agree in this aspect.

\section*{Acknowledgements}
We acknowledge Dr. Sen Wang, Profs. Volker Springel, Huiyuan Wang, Jing Wang, Yong Shi, Song Huang, and Xiaoyang Xia for their constructive and insightful suggestions. We also thank an anonymous referee for detailed and valuable comments, which have significantly improved the manuscript. This work is supported by the National Key Research and Development Program of China (grant No. 2022YFA1602903), the National Natural Science Foundation of China (grant No. 12433003), and the China Manned Space Project (No. CMS-CSST-2025-A10). This work acknowledges the Tsinghua Astrophysics High-Performance Computing platform for providing the computational and storage resources that supported this research. The authors gratefully acknowledge support from the Royal Society International Exchanges scheme (IES\textbackslash R2\textbackslash 242195). S.W. acknowledges support from China's National Foreign Expert program (H20240871).

%
\vspace{5mm}


\software{astropy \citep{astropy:2013, astropy:2018, astropy:2022},   
          Matplotlib \citep{Hunter:2007},
          NumPy \citep{numpy_harris2020array},
          SciPy \citep{2020SciPy-NMeth}
          }



\appendix
\section{Peak-valley pair searching algorithm}\label{sec:appendix_a}

Here, we present the peak-valley pair searching algorithm in detail. Given an SFR series of an individual galaxy, we first identify local maxima and minima by comparing each snapshot with its two immediate predecessors and two immediate successors (about 0.2 Gyr between adjacent snapshots), respectively. We then search for downward peak-valley pairs and upward valley-peak pairs that satisfy  $\Delta\log\,[\mathrm{SFR}/\mathrm{M_\odot Gyr^{-1}}]$ greater than a given threshold (e.g., 0.5 dex) since $z\sim 1$. 

Taking the upward valley-peak pair search for an example, the algorithm will be interpreted step by step:

\begin{enumerate}
    \item First, we start from the first minimum since $z\sim 1$ and label it as the valley candidate. 
    \item Then, we go forward in time for every minima and maxima in order and try to find a matching peak candidate that satisfies the threshold mentioned before. Several cases will be encountered during the process:
    \begin{enumerate}
        \item If, before finding an eligible peak candidate, another minimum lower than the initial minimum appears, we update the valley candidate to this lower one. 
        \item If a satisfied maximum is found, we label this maximum as the peak candidate. However, we will continue searching to see if there exists a maximum with an even greater difference. The purpose of this step is to dampen the effect of minor fluctuations and capture more significant signals. If so, we update the peak candidate to this new maximum. 
        \item The search process stops when the peak candidate exists and a further minimum is encountered such that the difference between this further minimum and the current peak candidate is comparable to that between the existing valley candidate and the current peak candidate (we set the ratio between the former and the latter to be larger than 0.7, i.e., $\mathrm{SFR}_{\rm peak}-\mathrm{SFR}_{\rm new\ min}>0.7\times(\mathrm{SFR}_{\rm peak}-\mathrm{SFR}_{\rm valley})$). Then, the current peak candidate and existing valley candidate will finally be recorded as an upward valley-peak pair. Setting this ratio to a value larger than 0.5 would not qualitatively change our results and conclusion. 
    \end{enumerate}
    \item Subsequently, the next starting valley will be selected as the first valley following the peak in the last pair, ensuring that pairs do not overlap.
\end{enumerate}

The procedure for identifying downward peak-valley pairs is analogous to that for the upward valley-peak pairs described above. For each galaxy, we search for both upward (i.e., valley-to-peak) pairs and downward (i.e., peak-to-valley) pairs. Among the total of 3644 star-forming central galaxies, 2536 galaxies ($\sim 70\%$) have experienced at least two upward or two downward pairs that satisfy  $\Delta\log\,[\mathrm{SFR}/\mathrm{M_\odot Gyr^{-1}}]>0.5$ dex since $z\sim 1$, which we consider to indicate the presence of at least two episodes.

\section{Comparison between the sample and the original distribution}\label{sec:appendix_samplecomparison}
\begin{figure*}[ht!]
    \centering
    \includegraphics[width=0.98\textwidth]{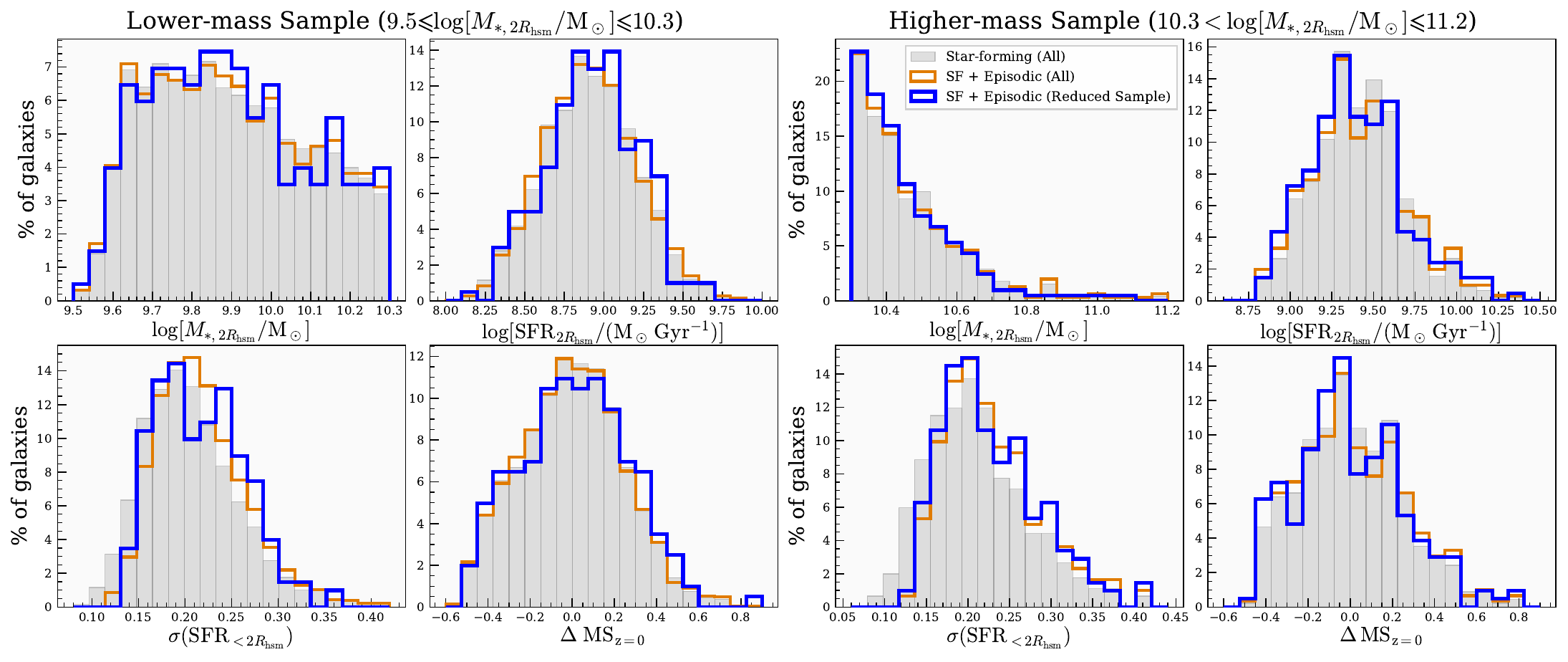}
    \caption{{\it Left panel:} the distribution of the central stellar mass $\log M_{\ast,2R_{\rm hsm}}$, SFR, present-day main-sequence offset $\Delta\rm MS_{z=0}$, and relative SFR fluctuation $\sigma_{\rm SFR, rel}$ of the lower-mass galaxy sample (blue), the total galaxies with typical episodic star-formation behavior (orange), and the total star-forming central galaxies (gray bars). {\it Right panel:} same as the left panel, but for the higher-mass galaxies.}
   \label{fig:samplecomparison}
\end{figure*}

In the left panel of Fig.\,\ref{fig:samplecomparison}, we present a comparison between the lower-mass galaxy sample, the total galaxies with typical episodic star-formation behavior, and the total star-forming central galaxies. The right panel compares the higher-mass galaxy sample with the latter two. The distributions are similar across the three samples, suggesting that our subsample does not introduce a systematic bias in the properties most relevant to our analysis.

\section{Different thresholds for peak-valley pairs}
\label{sec:appendix_threshold}

\begin{figure}[ht!]
    \centering
    \includegraphics[width=\linewidth]{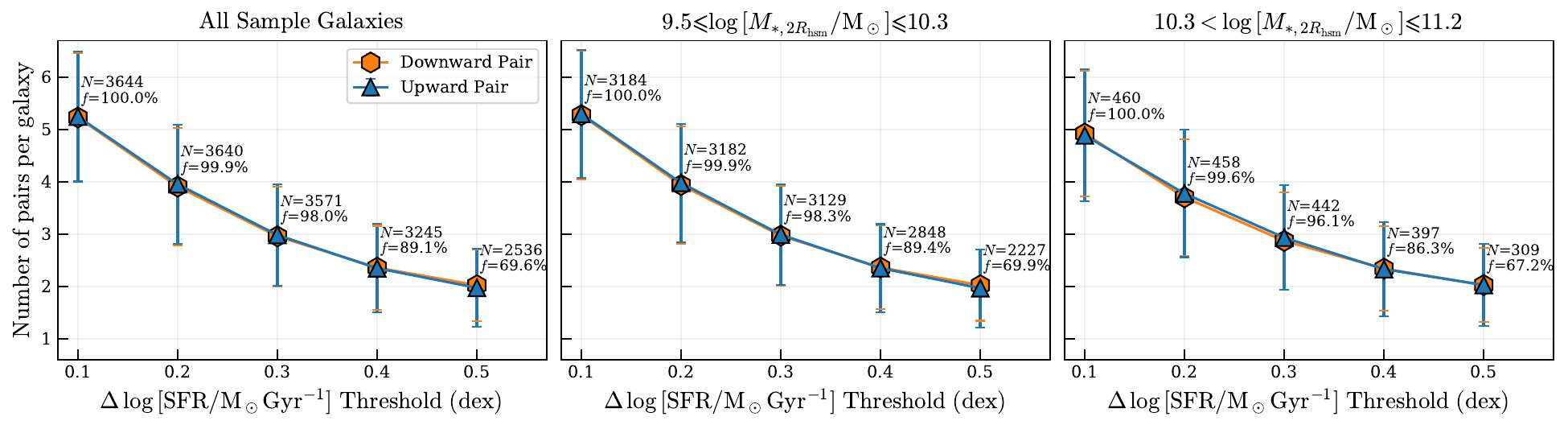}
    \caption{{\it Left panel:} the average number of pairs per galaxy as a function of $\Delta\log\,[\mathrm{SFR}/\mathrm{M_\odot Gyr^{-1}}]$ threshold, for star-forming central galaxies that have experienced at least two downward (orange hexagons) or upward (blue triangles) peak-valley pairs since $z=1$. The error bars indicate the standard deviations of the pair number. At each threshold, the number and fraction of galaxies that possess such episodic star-forming features are labeled beside the symbols. {\it Middle panel:} same as the left panel, but for the lower-mass galaxies with $9.5 \leqslant \log M_{\ast,2R_{\rm hsm}}/\mathrm{M_\odot} \leqslant 10.3$. {\it Right panel:} same as the left panel, but for the higher-mass galaxies with $10.3 < \log M_{\ast,2R_{\rm hsm}}/\mathrm{M_\odot} \leqslant 11.2$.}
    \label{fig:peakvalleynumber}
\end{figure}

\begin{figure}[ht!]
    \centering
    \includegraphics[width=\linewidth]{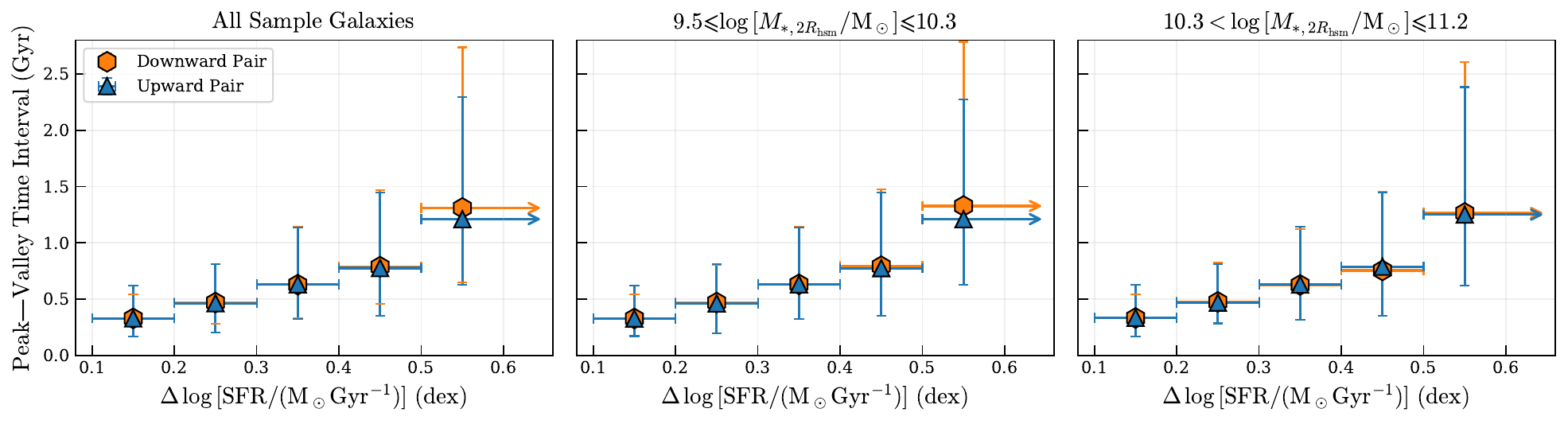}
    \caption{{\it Left panel:} the median and central 68th percentiles of time intervals of the downward (orange hexagons) and upward (blue triangles) pairs in different $\Delta\log\,[\mathrm{SFR}/\mathrm{M_\odot Gyr^{-1}}]$ bins with 0.1 dex interval, except for the last bin, where no upper bound is set for this bin. In this work, since we choose 0.5 dex as a threshold, the typical peak-valley timescale is $\sim$1.25 Gyr. {\it Middle panel:} same as the left panel, but for the lower-mass galaxies with $9.5 \leqslant \log M_{\ast,2R_{\rm hsm}}/\mathrm{M_\odot} \leqslant 10.3$. {\it Right panel:} same as the left panel, but for the higher-mass galaxies with $10.3 < \log M_{\ast,2R_{\rm hsm}}/\mathrm{M_\odot} \leqslant 11.2$.}
    \label{fig:timeintervals}
\end{figure}

In this study, we only take galaxies that possess at least two peak-valley pairs with $\Delta\log\,[\mathrm{SFR}/\mathrm{M_\odot Gyr^{-1}}]>0.5$ dex to further compose a refined galaxy sample for peak-valley statistics (see Section \ref{sec:method_SelESFG} for details). We note that, however, the number of peak-valley pairs per galaxy (for typical episodic star-forming galaxies that possess at least two upward or downward pairs) and the time interval in between depend on $\Delta\log\,[\mathrm{SFR}/\mathrm{M_\odot Gyr^{-1}}]$.

We investigated five different $\Delta\log\,[\mathrm{SFR}/\mathrm{M_\odot Gyr^{-1}}]$ thresholds, being 0.1, 0.2, 0.3, 0.4, and 0.5 dex. The corresponding results are presented in Fig.\,\ref{fig:peakvalleynumber} and Fig.\,\ref{fig:timeintervals}. Generally speaking, a larger (smaller) fluctuation amplitude (i.e., $\Delta\log\,[\mathrm{SFR}/\mathrm{M_\odot Gyr^{-1}}]$) leads to fewer (more) peak-valley pairs with longer (shorter) timescales. As can be seen, with $\Delta\log\,[\mathrm{SFR}/\mathrm{M_\odot Gyr^{-1}}]\geqslant $ 0.3 dex, nearly 98\% of the 3644 star-forming central galaxies would on average have three peak-valley pairs since $z=1$, with an average time interval between them $\sim 600$ Myr. As the $\Delta\log\,[\mathrm{SFR}/\mathrm{M_\odot Gyr^{-1}}]$ threshold increases to 0.5 dex, $\sim 70\%$ (i.e., 2536) of the full sample have on average two peak-valley pairs with an average time interval of $\sim$1.25 Gyr. We further examine the results in two mass bins. As can be seen, the number of pairs per galaxy and the peak-valley time intervals are essentially the same between galaxies in the lower- and higher-mass bins.

We note that through visual inspection over all 3644 star-forming central galaxies. we confirm that the reported two branches of star formation with a retreating/shrinking disk growing fashion are universal among this population, regardless of the $\Delta\log\,[\mathrm{SFR}/\mathrm{M_\odot Gyr^{-1}}]$ threshold adopted (see Section \ref{sec:results-overview} for details).

\section{Stacking using absolute time intervals}\label{sec:appendix_stacking}
\begin{figure*}[ht!]
    \flushleft
    \includegraphics[width=0.98\textwidth]{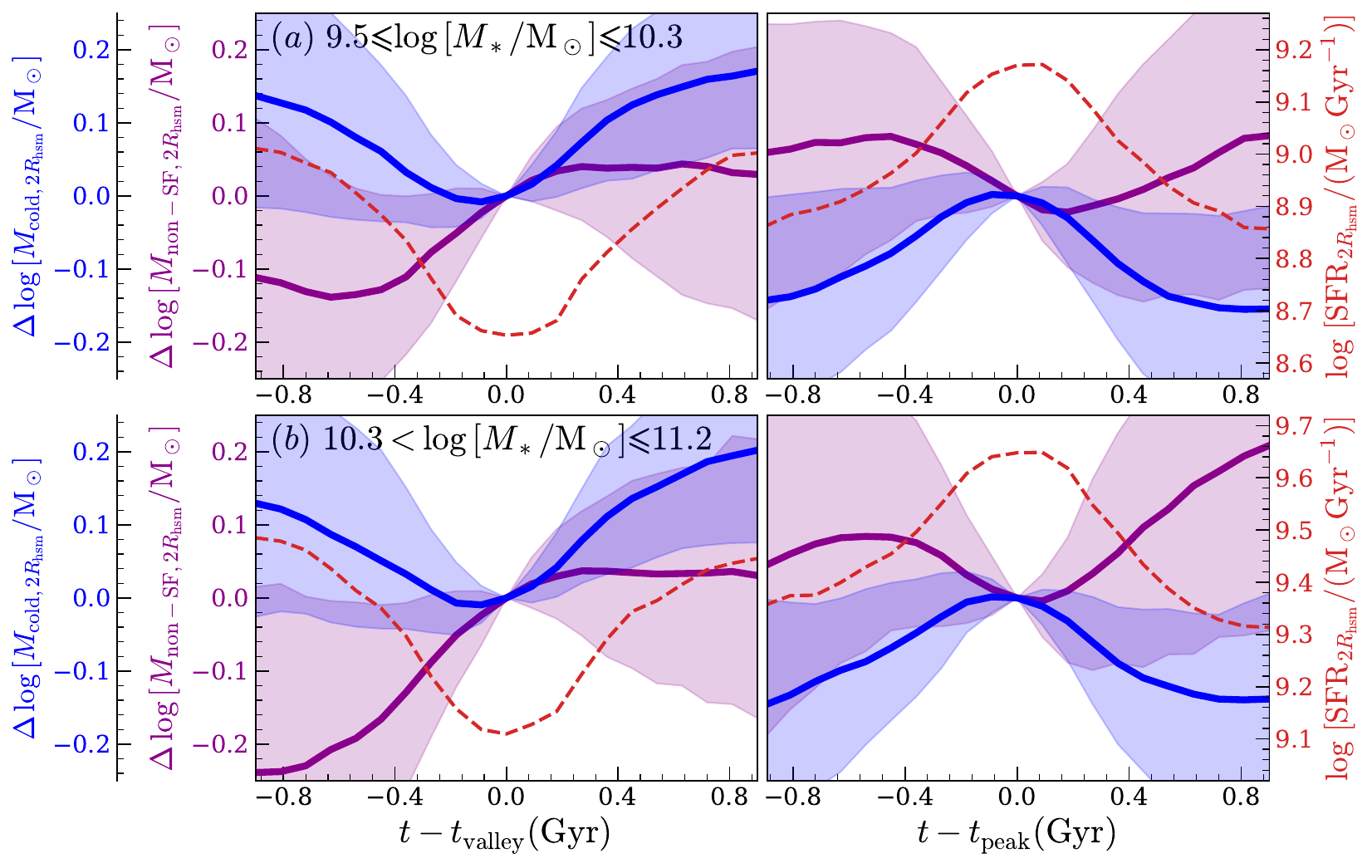}
    \caption{Similar to Fig.\,\ref{fig:gasZ_nearpeva_seperated}, but stacking the relative cold gas mass and cold but {\it non-SF} gas as a function of absolute time offset (in Gyr) around the nearest peak/valley moment.}
   \label{fig:gasZ_nearpeva_seperated_abstime}
\end{figure*}

For comparison with Fig.\,\ref{fig:gasZ_nearpeva_seperated}, we stack the cold gas within $2R_{\mathrm{hsm}}$ relative to its values at peaks and valleys as a function of absolute time offset around these critical moments, which is shown in Fig.\,\ref{fig:gasZ_nearpeva_seperated_abstime}. The results are qualitatively consistent with Fig.\,\ref{fig:gasZ_nearpeva_seperated}.


\bibliography{EpisodicSFH}
\bibliographystyle{aasjournal}



\end{document}